**Integrating audiological datasets via federated merging of Auditory Profiles**

Samira Saak[1,2], Dirk Oetting[2,3], Birger Kollmeier[1,2,3], Mareike Buhl[2,4]

[1] Medizinische Physik, Carl von Ossietzky Universität Oldenburg, Oldenburg, Germany
[2] Cluster of Excellence „Hearing4all", Carl von Ossietzky Universität Oldenburg, Oldenburg, Germany
[3] Hörzentrum Oldenburg gGmbH, Oldenburg, Germany
[4] Université Paris Cité, Institut Pasteur, AP-HP, Inserm, Fondation Pour l'Audition, Institut de l'Audition, IHU reConnect, F-75012 Paris, France

---

Audiological datasets contain valuable knowledge about hearing loss in patients, which can be uncovered using data-driven, federated learning techniques. Our previous approach summarized patient information from one audiological dataset into distinct Auditory Profiles (APs). To obtain a better estimate of the audiological patient population, however, patient patterns must be analyzed across multiple, separated datasets, and finally, be integrated into a combined set of APs.

This study aimed at extending the existing profile generation pipeline with an AP merging step, enabling the combination of APs from different datasets based on their similarity across audiological measures. The 13 previously generated APs ($N_A$=595) were merged with 31 newly generated APs from a second dataset ($N_B$=1272) using a similarity score derived from the overlapping densities of common features across the two datasets. To ensure clinical applicability, random forest models were created for various scenarios, encompassing different combinations of audiological measures.

A new set with 13 combined APs is proposed, providing separable profiles, which still capture detailed patient information from various test outcome combinations. The classification performance across these profiles is satisfactory. The best performance was achieved using a combination of loudness scaling, audiogram and speech test information, while single measures performed worst.

The enhanced profile generation pipeline demonstrates the feasibility of combining APs across datasets, which should generalize to all datasets and could lead to an interpretable global profile set in the future. The classification models maintain clinical applicability.

---

Corresponding author: Samira Saak, samira.kristina.saak@uol.de

## I. Introduction

Audiological datasets contain valuable knowledge about patients with hearing loss that can be exploited to learn about patterns in the data, for instance, for identifying patient groups that exhibit similar combinations of audiological test outcomes and may therefore benefit from a similar treatment. Data-driven techniques allow assessing these feature combinations without prior knowledge about audiological findings or diagnostic information. Performing such analyses on large-scale datasets has received increasing attention across medical fields as well as by policy-makers, due to its potential for precision medicine, the identification of risk factors for diseases, and health care quality control ((Sinkala et al., 2020; Stöver et al., 2023), (European Commission. Directorate General for Health and Food Safety. et al., 2016), to name a few). In the domain of audiology, available data for large-scale analyses is currently spread across institutions. While research institutes often have smaller datasets available and may share them openly, larger datasets are available in clinics, but these are restricted in terms of access. Further, clinical or research institutions generally obtain data from patient groups in line with the respective institution's purpose or purpose of the respective study which may lead to considerably different patient populations and audiological measures across data sets. For example, a cochlear implant clinic will have a larger proportion of severe hearing deficits in its datasets than an ambulant audiological center, which will be frequented more often by hearing aid candidates. Even more variability across datasets comes into play if online remote testing with smartphones is included, allowing for data collection outside labs and clinics with a population exhibiting a mild-to-moderate hearing loss. Hence, a comprehensive data analysis describing the complete audiological patient population requires the combination of available, but distributed data across institutions without having the access to all databases simultaneously.

Federated learning has been shown to overcome the problem of distributed data ownership (Konečný et al., 2017; McMahan et al., 2017; Pfitzner et al., 2021). By training machine learning models locally at the data ownership site, federated learning enables the integration of sensitive data, including clinical datasets that were previously inaccessible due to data restrictions (European Parliament & European Council, 2016). Only anonymized updates from these local models are shared and aggregated with a global model, which is then used to improve each local model. This iterative process updates the global model to improve performance. However, this requires frequent communication between local and global models, which may not always be feasible, particularly in research collaborations involving medical institutions. In such cases, one-shot federated learning may be more suitable, as it only requires a single communication round (Duan et al., 2024; Guha et al., 2019). One-shot federated learning involves training models locally and then aggregating the distilled knowledge into a global model. In line with one-shot federated learning, this study introduces a framework for federated merging of Auditory Profiles by demonstrating the feasibility of sequentially extracting information from two different databases and then merging the resulting information in a second step.

Varying approaches exist to describe the audiological patient population. These approaches range from epidemiological studies regarding the prevalence of hearing loss in combination with different demographic factors (do Carmo et al., 2008; Roth et al., 2011), to developing optimal test batteries to characterize hearing deficits (Sanchez-Lopez et al., 2021; van Esch et al., 2013), and advanced analyses into audiological groupings (Bisgaard et al., 2010; Saak et al., 2022; Sanchez-Lopez et al., 2020). Bisgaard et al. (2010) grouped audiograms into ten standard audiogram patterns. However, as no additional measures are considered, a detailed description based on multiple domains of hearing, including, e.g., suprathreshold auditory processing deficits like Plomp's (Plomp, 1986) "D-component", is not possible. Sanchez-Lopez et al. (2020) applied principal component analysis and archetypal analyses to multiple audiological measures to characterize patients into four distinct auditory profiles. By defining the auditory profiles based on the hypothesis of two underlying distortion types they limit the number of profiles to four. Saak et al. (2022) proposed a profile generation pipeline which resulted in a first set of 13 auditory profiles (APs). Again, APs were generated based on multiple audiological measures (e.g. speech testing, loudness scaling, audiogram). In contrast to Sanchez-Lopez et al. (2020), the number of APs is not chosen based on a hypothesis, but dependent on the patient population of the underlying dataset, hence, completely data-driven. APs are described in terms of data distributions across measures, which has the potential to integrate additional profiles into a combined set of APs. If sufficient datasets are accounted for that cover a large variety of the audiological patient population, such a combined set of APs could ultimately lead to a robust, comprehensive AP set that can provide a better estimate of the audiological patient population than single datasets.

To create such a comprehensive AP set, it is necessary to compare and integrate multiple datasets from various sources, leveraging the growing amount of available data, including various subtypes of hearing deficits as well as different information collected in the respective dataset. Combining multiple datasets from diverse institutions and geographical locations enables APs to more accurately reflect the audiological patient population. This is because multiple datasets can provide a more unbiased estimate than a single dataset, unless patients were acquired via stratified sampling. Hence, a suitable tool must be capable of integrating APs from different datasets including different tests and features, while ensuring interpretability, flexibility, applicability to clinical practice, as well as complying with data privacy and protection regulations by implementing the federated learning principle (Konečný et al., 2017; McMahan et al., 2017).

First, the process to combine profiles generated across datasets, as well as the resulting profiles need to remain interpretable to ensure the plausibility and applicability of the profiles in clinical practice. In clinical practice, for instance, if patients are classified into the APs, the APs could serve as a look-up table to obtain an estimation of the results on further audiological tests included in the AP but not measured by the practitioner. Only if their generation is interpretable and the resulting combined profiles are plausible, valuable insights can be obtained from these profiles. Finding such interpretable subgroups or subtypes of diseases via big data analyses has proven valuable and transformative for medical research, and precision medicine (Grant et al., 2020; Parimbelli et al., 2018). For example, molecular profiling has allowed to identify two pancreatic cancer subtypes with subtype-specific biomarkers. The two subtypes have different clinical outcomes and may be more responsive to different subtype-specific treatments (Sinkala et al., 2020). In a similar way, APs could serve to identify subgroup-specific causes and treatment recommendations, for instance, by highlighting patient groups that exhibit a similar audiogram but may benefit from different fitting rationales for the hearing aids or by prompting sub-group specific research.

Second, it is important to achieve a flexible process for combining APs, such that they can be used in a variety of settings. For instance, for research purposes certain distinctions across audiological measures may be of interest, whereas for clinical applications a coarser separation may suffice. More detailed profiles would represent subtypes of coarser profiles and could, thus, inform on potential subgroups of individual profiles. Hence, providing a method that can make profiles more detailed or coarser based on the individual use case helps in the applicability of the profiles in practice.

Third, for APs to be useful in audiological/clinical practice, the identified subgroup-specific causes and treatment recommendations need to be accessible to practitioners, such as clinicians, hearing care professionals, and researchers. Hence, different classification models are required that are based on different subsets of audiological measurements (features) corresponding to the needs of the respective clinicians or researchers. For example, hearing care professionals in Germany, thus far, mainly use the audiogram and the Freiburger speech test in quiet in their hearing aid provision process, as these are required for a hearing aid indication and hearing aid approval for the reimbursement of the health care insurance (Gemeinsamer Bundesausschuss, 2021), even though research indicates that loudness scaling can be beneficial in hearing aid fitting (Oetting et al, 2018). Smartphone apps, in contrast, can easily perform loudness scaling (Xu et al., 2024), but measuring bone conduction thresholds without special equipment is not as straightforward, even though conductive hearing loss can be estimated with the antiphasic digits-in-noise test (DIN) (De Sousa et al., 2022).

Finally, the process to combine and compare profiles needs to be able to also handle sensitive data, such that clinical datasets can be used that underlie data restrictions (European Parliament & European Council, 2016). Federated learning, as proposed by Szatmari et al. (2020) for modelling audiological preferences, offers a promising solution for training machine learning models remotely while adhering to privacy constraints. However, when integrating additional datasets for auditory profiles, multiple rounds of updates may not be feasible. In such cases, a one-shot federated learning framework (McMahan et al., 2017; Pfitzner et al., 2021) could serve as a solution for data privacy restrictions. That means, local APs could be generated at a specific institution, and the AP information could then be shared to update the already existing global APs. In other words, insights and knowledge could be extracted from different datasets via the APs, without the need to combine the respective datasets.

The present study aims to develop a flexible and interpretable approach for combining auditory profiles (APs), generated with the analysis pipeline of Saak et al. (2022), that complies with data protection standards and can be used in clinical practice. That means, we aim to investigate how APs can be merged from different datasets in a federated way to allow for knowledge integration via APs (while maintaining information from all tests) and take the next steps towards an audiological patient population-based estimate of APs. To achieve this, we (1) aim to compare APs generated on two different datasets with respect to their profile similarity. To ensure the interpretability of the merging approach, we assess both the overall and feature-specific profile similarity across all merging iterations. We hypothesize that some APs will be similar across datasets, while the inclusion of further datasets will also result in new AP patterns, as a broader range of audiological patients is covered. Further, we (2) aim to analyze the feature importance of the distinct APs, such that APs can be easily identified by their specific patterns across audiological measures. Finally, we (3) aim to make the APs applicable by providing classification models for various settings, including research, hearing aid fitting, and potential smartphone applications. The current study, thus, aims to answer the following research questions:

**RQ1:** How can we obtain a combined set of auditory profiles from two audiological datasets?

**RQ1.1:** How can auditory profiles, generated on different datasets, be merged, such that a privacy-preserving combination of profiles from different datasets can be achieved?

**RQ1.2:** Are the previously generated 13 auditory profiles also represented in the second audiological dataset?

**RQ1.3:** How can we ensure that the proposed merging approach can be flexibly used in different use cases and how does flexibility relate to feature importance in different merging steps?

**RQ2:** How well can we classify patients into the generated profiles and which features are most important for this?

**RQ3:** Are the audiological features that are deemed important in the merging iterations also important for classifying patients into the auditory profiles?

## II. Method

### II.1 Datasets

We used two separate datasets (dataset A and B) in our analysis to generate two separate AP sets (profile set A and B). Both datasets were provided by the Hörzentrum Oldenburg gGmbH (Germany) and contain information with respect to a broad range of measures. They include measures contained in both datasets, which we refer to as common measures, as well as different additional measures. Common measures for both datasets are age, the Goettingen Sentence Test (GOESA, Kollmeier & Wesselkamp (1997)), the audiogram for air- and bone-conduction, and the Adaptive Categorical Loudness Scaling (ACALOS, Brand & Hohmann (2002)). Both datasets use narrowband noise for ACALOS for the left and right ear measured with headphones. For dataset A results for the center frequencies 1.5 and 4 kHz are available and for dataset B for 1 and 4 kHz. For audiogram and ACALOS measures, we only used the worse ear to reduce the feature space, but account for asymmetric hearing loss by introducing an asymmetry score derived from the audiogram (ASYM). In the following, we will refer to audiological tests as *measures,* and to *features* both as specific sub-results of the measures and as summarizing features.

For example, a common feature between the two datasets of the measure GOESA is the binaural speech recognition threshold (SRT) for the collocated S0N0 condition, measured adaptively with varying speech level and fixed test-specific noise level of 65 dB SPL (specific sub-result). For the audiogram, the common features we used are summarizing features, derived from specific sub-results. These include the pure tone average (PTA, calculated from the hearing threshold at 0.5, 1, 2, and 4 kHz) for air (AC PTA) and bone (BC PTA) conduction, an asymmetry score between left and right ear (difference between left and right PTAs, ASYM), the air-bone-gap (ABG), the Bisgaard class (Bisgaard et al., 2010) to characterize the shape of the audiogram, and the PTA of the uncomfortable level (UCL PTA). For ACALOS we used summarizing features that can characterize both the lower and upper part of the loudness curve ($L_{15}$, $L_{35}$, i.e., level corresponding to categorical loudness of 15 CU and 35 CU, respectively), and the difference between $L_{15}$ and $L_{35}$ (as a feature representing the dynamic range), for both available frequencies (i.e., 1_diff and 4_diff). More information on common features between the two datasets and feature combinations used for analysis is given in Table 1. A description of ranges across common and additional features of the two datasets can be found in the supplementary material.

#### II.1.1 Original dataset A

The original dataset A refers to the dataset and the profile set used in Saak et al. (2022). The dataset was collected for research purposes (Gieseler et al., 2017) and consists of 595 listeners (*mean age* = 67.6, *SD* = 11.9, *female* = 44%) with normal to impaired hearing. Next to the common measures and features, additional information is contained. In the speech test domain, the SRT for the digit triplet test (Zokoll et al., 2012) and the slope for the GOESA S0N0 condition are available. Further, information regarding the socio-economic status and two cognitive measures are included, namely the Demtect (Kalbe et al., 2004) and the Vocabulary test (Schmidt & Metzler, 1992). All these measures were used for the generation of the profiles. Detailed information about this dataset can be found in Gieseler et al. (2017) and Saak et al. (2022).

#### II.1.2 Dataset B

The new dataset B was collected for diagnostic purposes and consists of 1401 listeners. The main measures overlap with dataset A (common measures). However, no cognitive measures are available for dataset B. Further, additional GOESA conditions are available, namely the S0N90 binaural and monaural conditions, where noise is presented from one side. Due to the diagnostic purpose of the dataset, information on the tympanogram (Type A, As, Ad, B, C, D, tympanic membrane perforation, not measurable), Valsalva and otoscopy (not impaired, moderately impaired, impaired) is also available for both ears. All these measures were used for the generation of the profiles. Only patients with data for at least two common measures (from the audiogram, speech test, and loudness scaling) were included to ensure sufficient information for the clustering. This resulted in 1272 patients (*mean age* = 63.74, *SD* = 13.22*, female* = 42.26%) with normal to impaired hearing. In addition, different outcome parameters are available, such as International Statistical Classification of Diseases and related Health Problems (ICD) codes, and information on a potential hearing aid supply and the respective aided performance. Outcome parameters were not used for the generation of the profiles.

### II.2 One-shot federated learning framework for generating auditory profiles from multiple datasets

To enable the integration of sensitive data that underly privacy restrictions we extended the previously developed profiling pipeline by Saak et al., (Saak et al., 2022) to a one-shot federated learning framework for data-driven characterization of audiological databases. Using a one-shot federated learning framework enables us to only require one communication round between data ownership site and central computation. Specifically, this means that local profiles can be generated locally at the data ownership site. These profiles represent distilled knowledge from the datasets that are then shared to construct the global profile set. This process is described in section II.2.2. Figure 1 visualizes the one-shot federated learning framework for the auditory profiles.

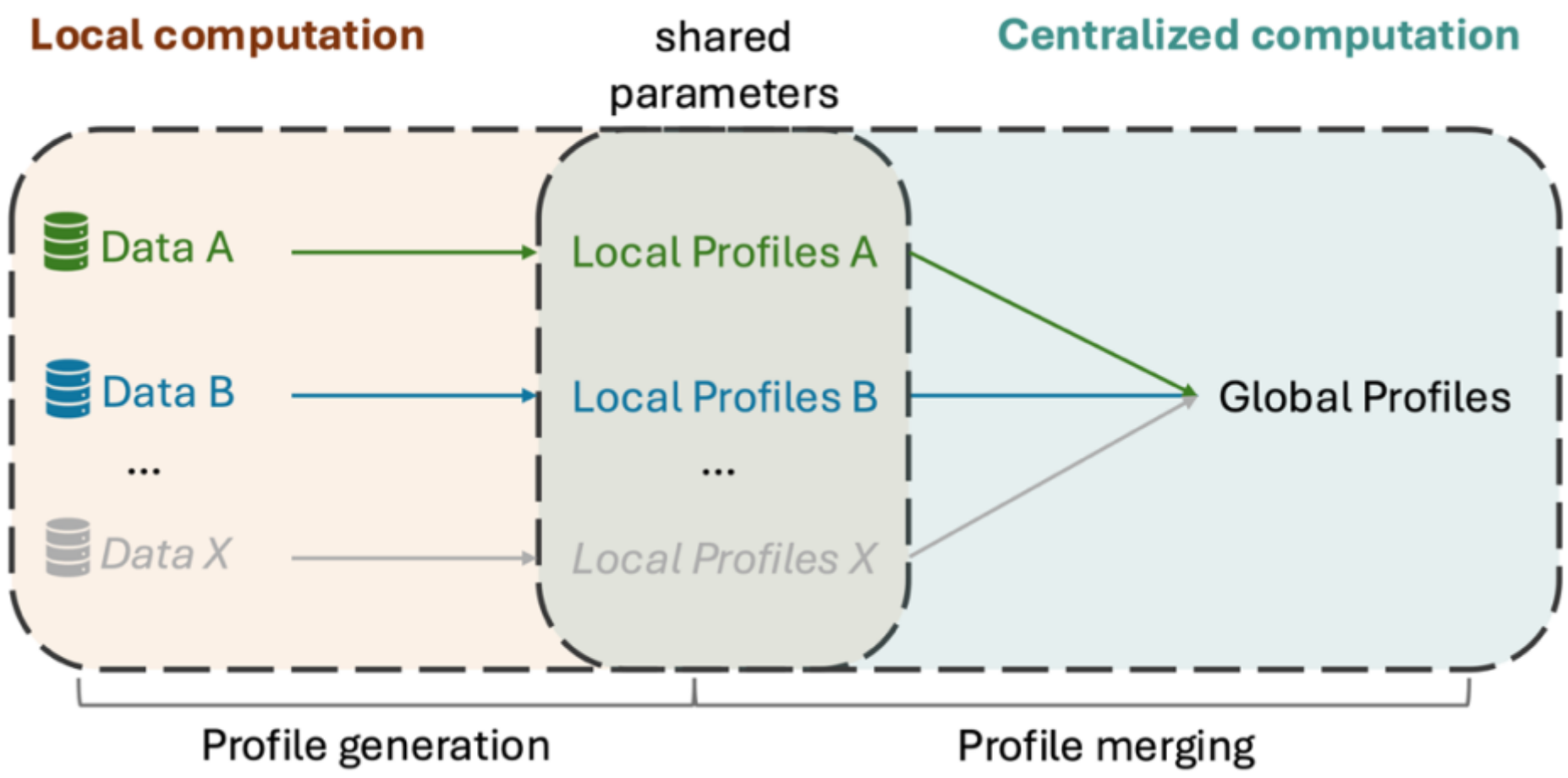

*Figure 1: Visualization of the one-shot federated learning framework for auditory profiles. Profile generation can occur locally, while profile merging is performed centrally. Grey italic text shows the potential addition of further datasets.*

### II.2.1 Profile generation

For dataset A, APs were already available and obtained from (Saak et al., 2022). They contain 13 distinct APs, and will be referred to as profile set A. Each profile consists of distributions across different features from audiological measures (audiogram, ACALOS, GOESA, …). These ranges provide an estimate of plausible values for individuals that belong to a certain profile. To generate the profiles, both the common and additional features from dataset A were used.

For dataset B, APs (profile set B) were generated using the profile generation pipeline from (Saak et al., 2022). Features used for profile generation include the common and additional features from dataset B. No outcome measures, such as the ICD codes, were included to cluster patients into the APs. The available categorical features from the tympanogram, otoscopy, and Valsalva were transformed to continuous features using multiple correspondence analysis (MCA, (Lê et al., 2008). MCA is a dimension reduction method, similar to principal component analyses. It quantifies the relationship between categorical variables in the form of principal components, and in that way transforms the categorical features into continuous components. We used the first three resulting components instead of the original three categorical variables as features for the subsequent data analyses. With this approach we retained some information regarding the three categorical variables for the generation of profiles. The resulting components, however, only represent the relationship across the categorical variables and some information will be missing. Hence, in the future a different approach to tackle the difficulty of handling mixed data may become preferable.

Model-based clustering was used to generate the profiles. Model-based clustering assumes that an underlying model generated the data, and the clustering aims at recovering the model (Banerjee & Shan, 2017). The model is a combination of data clusters that describe the patterns in the data. These clusters serve as the generated APs. The profile generation pipeline (Saak et al., 2022) using model-based clustering consists of two steps:

In the robust learning step two parameters are estimated, namely the number of profiles present in the dataset and the respective covariance parameterization. To achieve this, bootstrapping without replacement was used to generate 1000 datasets, each using 90% of the data. Next, missing data in each bootstrapped dataset was imputed. Here, we made a small adjustment to the original pipeline, which is expected to not impact the clustering results, as internal simulation analyses have confirmed for both datasets. We replaced Multivariate Imputation by Chained Equations (MICE, (Buuren & Groothuis-Oudshoorn, 2011) by imputation based on factorial analysis for mixed data (FAMD, (Audigier et al., 2016). FAMD is a competitive imputation technique based on principal components that reduces the computational complexity of the profile generation pipeline. This means, that instead of multiple imputed datasets only a single imputed dataset is generated for each bootstrapped dataset, which simplifies the following computations for cluster analyses. Next, features were scaled using the min-max scaling, which transforms values to range from 0 to 1. We then applied model-based clustering to each dataset using different parameter combinations (number of profiles, covariance parameterization). The possible number of profiles was set to 1 to 40 profiles to cover a broad potential range of profiles. The covariance parameterizations determine the shape, volume and orientation of the profiles (see (Fraley & Raftery, 2003) for all possible covariance parameterizations). The model describing the underlying data best was then selected using the Bayesian information criterion (Schwarz, 1978). Finally, the most frequently occurring number of profiles and covariance matrix across all bootstrapped datasets was selected as the model parameter solution fitting the data best.

The profile generation step uses the original complete dataset and, again, imputes missing data with FAMD. Model-based clustering is then applied to the scaled features with the learned model parameter solution to result in the APs of profile set B. To allow for later merging with anonymized data distributions from sensitive datasets, all profile data was transformed to count data with 100 equidistant steps. More details regarding the generation of profiles can be found in (Saak et al., 2022).

### II.2.2 Profile merging

To compare and combine profiles generated across multiple datasets, they need to share common features that allow to investigate how similar different profiles are. We, therefore, selected the common features of the two datasets across the domains of speech intelligibility, audiogram, loudness scaling, and anamnesis information, namely the age of the individual. The rows of Table 1 display all features used to combine the two datasets.

#### II.2.2.1 Overlapping density index

A measure of similarity is required that allows for an estimation of similarity between two profiles. For this purpose, we use the overlapping index (Pastore & Calcagnì, 2019). The overlapping index is a distribution-free index that calculates the overlapping area of two probability density functions. To obtain the similarity between two respective profiles we calculate the overlapping index for each feature of the two profiles and use the mean as the final similarity score for the respective profiles. Sample size imbalances are accounted for by using normalized density distributions.

#### II.2.2.2 Merging procedure

Starting from all profiles generated on datasets A and B (44 profiles, combination of profiles estimated separately for dataset A and B, respectively), profiles were merged iteratively, and the procedure is depicted in Figure 2. The similarity score was calculated for all profile combinations, and the two profiles with the highest similarity score were merged. The SRT was used 6 times when calculating the similarity score to balance the effect of the speech test to the number of available features from the audiogram and ACALOS. Then, for the new set of profiles containing the merged profile, the similarity score was calculated for all profile combinations, and the two profiles with the highest similarity score were merged. This procedure continued until only two profiles remained.

Hence, we pre-generated all potential profile solutions and then derived a stopping criterion to result in the final profile set, as described in section II.2.2.3.

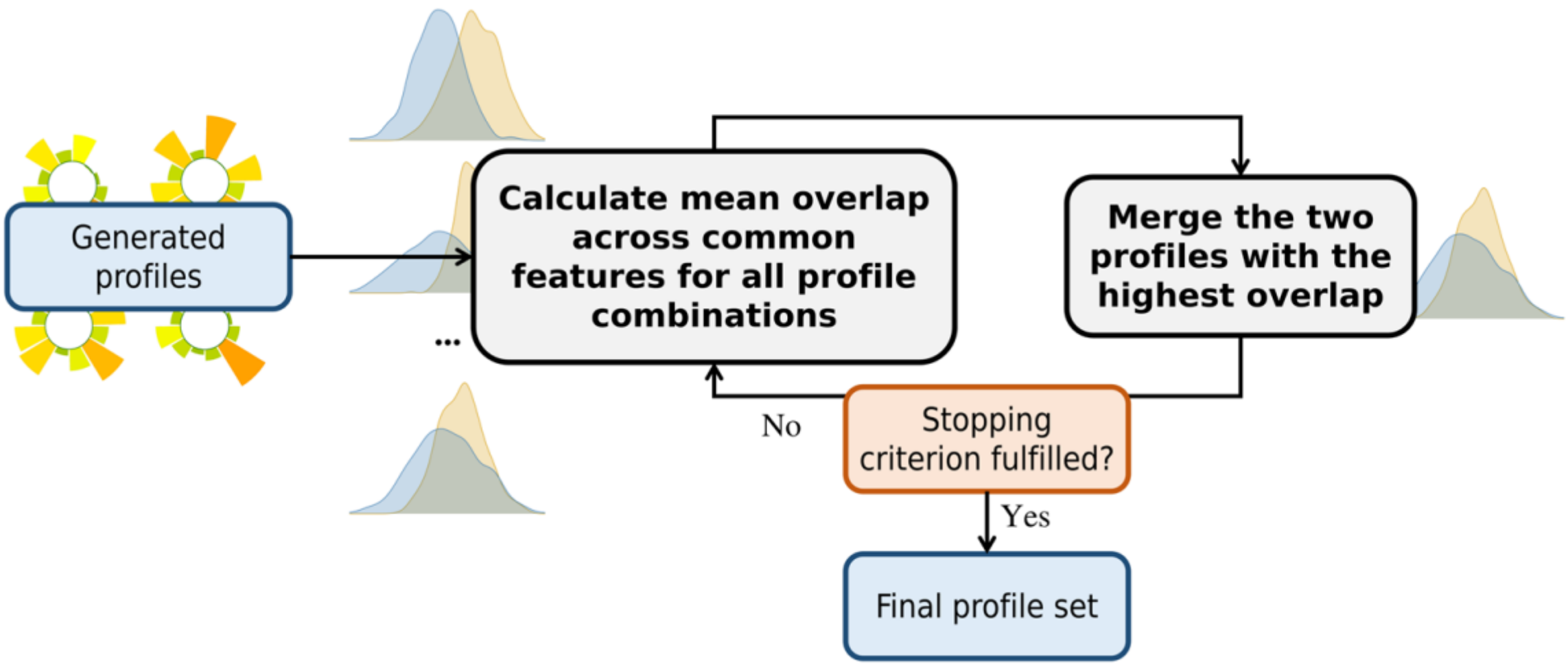

*Figure 2: Merging procedure. Profiles are merged iteratively until a stopping criterion is fulfilled. The two profiles with the highest overlap are merged at each iteration step*

### II.2.2.3 Selection of merged profile set

After each merge a potential profile set is generated. That means the merging pipeline results in *N* - 1 potential profile sets, ranging from all *N* available profiles from the two datasets to the last two remaining profiles. Depending on the intended use case in practice, a plausible profile set needs to be selected. For instance, if a coarse classification into mild, moderate, and severe hearing loss is desired, merging could continue until only two to three profiles remain, instead of stopping the merging iteration earlier. This does not, however, provide a detailed differentiation across features, which can be achieved with a larger profile set, potentially necessary for research purposes or detailed patient characterization. In contrast, a profile set containing a larger number of profiles may contain redundant information for clinical classification but may aid in investigating specific differences between patient groups. For that reason, we aimed at proposing a basic set of APs that remains detailed enough, whilst reducing the number of profiles and combining similar profiles contained in both datasets. More specifically, we aimed at reducing redundancy across profiles for a proposed general set of profiles, such that the differences between the profiles are maximized. To achieve this, we used a combination of two steps to manually select a cutoff based on the overlapping density (see Figure 4A in the results section), corresponding to a number of obtained profiles at a certain number of merging iterations.

In the first step, we selected the cutoff based on two parameters, namely the slope of the maximum overlapping density, and the variance of the median overlapping density. The first parameter describes the slope over merging iterations for the two profiles with the highest overlap in each merging iteration. A steeper slope and lower overlap, thus, indicates when profiles are merged that are less similar than in previous steps. We selected the parameter such that the slope decrease is relatively larger after the cutoff, as compared to prior to the cutoff. The second parameter describes the variance over iterations for the median overlap of all profiles. If the variance changes too much over iterations, it indicates that merges took place between two profiles that differ strongly from each other. Hence, we selected the cutoff such that the variance remains relatively stable prior to the cutoff, contrasting the variance after the cutoff.

In the second step, we compared the overlapping index across features before and after the previous cutoff. For an optimal cutoff, we expect a higher overlapping index prior to the cutoff, and a lower overlapping index after the cutoff. In that way, we can determine whether features were overall similar for the two profiles to be merged, or whether they differ substantially from each other, in which case a merge may not be advisable. Hence, it also allows to investigate which features are most important for the merging procedure and the profiles, and likewise, which features distinguish two profiles the most. We aimed for a cutoff that shows high overlap of features prior to it, and substantial difference between profiles after the cutoff.

#### II.2.2.4 Feature importance of the merges

For an interpretable merging procedure and interpretable APs, it is highly relevant to investigate which audiological measures drive the profile merges (high overlap), and which audiological measures hinder the profile merges (low overlap). That is because features that drive the profile merges are less relevant, whereas features that hinder the profile merges are more relevant and able to discriminate better between profiles. We, therefore, investigated which features had the least overlap across profiles, that is, were most discriminative between profiles. For that purpose, we defined different sets of iterations, for instance, iterations before the defined cutoff and after the defined cutoff. Iterations before the defined cutoff indicate which features were merged and which information may have been lost. Iterations after the defined cutoff indicate which features are most important for discriminating between profiles. For example, if we start at the last profile set with two remaining profiles, we can investigate which measures are least similar across the two profiles and would therefore lead to a split into three profiles. Hence, these features would support the split and can be considered as relevant features.

We further investigated which features were most relevant across different ranges of iterations (merging areas, see A-E in Figure 4B). For this, we calculated the mean overlap for each feature within a merging area and compared this to the average mean overlap for all features for the respective merging area. Hence, we investigated whether a feature was more or less important than the average of all features within the merging area.

### II.3 Classification models

We built classification models for the global APs for two reasons. First, we aimed at enabling the classification of new patients into the profiles, such that the profiles can be used in practice. Second, the feature importance of the classification models allows to draw conclusions with respect to which features are most relevant for classification into the respective profiles.

#### II.3.1 Feature sets and labels

To simulate different use cases, enable applicability of the profiles in practice, and extract insights into feature importance, different feature sets (subsets of the common features used for merging) were used to build the classification models. The feature sets belong to three general categories, denotes as use cases, combined, and single in the following. The feature combinations "ALL", "APP" and "HA" belong to the category use case, as they represent specific use cases in practice. "ALL" contains all common features, enabling a comprehensive feature importance interpretation across all features. "APP" defines features that could potentially be measured via a smartphone, facilitating the estimation of APs in a mobile context. "HA" defines features generally available for hearing care professionals, making it possible to use APs in a hearing aid fitting context. The combined feature group explores the performance of only using two of the three main features, while the single feature group investigates performance with each measure separately. In that way, the classification models could be used also for datasets only containing one or two of the three measures. Specifically, "AG" includes audiogram-related features, "SRT" includes speech test-related features, and "ACALOS" includes ACALOS-related features. Consequently, "AG SRT", "AG ACALOS", and "SRT ACALOS" represent the combination of these respective tests. All sets are displayed in Table 1. While these feature groups allow for broader applicability of the classification models, they also allow for feature importance interpretations, as the performance across feature sets can be compared. Profile numbers of the final combined profile set were used as labels for classification.

*Table 1: The different feature sets used in the analysis (see II.1 for feature descriptions). "ALL" corresponds to all features common to both datasets that were also used for the profile generation; "APP" to measures potentially measurable via a smartphone; "HA" to measures generally available for hearing care professionals. The remaining feature groups describe feature combinations and the importance of single features.*

| | | **Use cases** | | | **Combined** | | | **Single** | | |
|---|---|---|---|---|---|---|---|---|---|---|
| | | **ALL** | **APP** | **HA** | **AG SRT** | **AG ACALOS** | **SRT ACALOS** | **AG** | **SRT** | **ACALOS** |
| **Speech test** | S0N0 bin | ✓ | ✓ | ✓ | ✓ | | ✓ | | ✓ | |
| **Audiogram** | AC PTA | ✓ | ✓ | ✓ | ✓ | ✓ | | ✓ | | |
| | ASYM | ✓ | ✓ | ✓ | ✓ | ✓ | | ✓ | | |
| | BISGAARD | ✓ | ✓ | ✓ | ✓ | ✓ | | ✓ | | |
| | BC PTA | ✓ | | ✓ | | | | | | |
| | ABG | ✓ | | ✓ | | | | | | |
| | UCL PTA | ✓ | | ✓ | | | | | | |
| **Anamnesis** | Age | ✓ | ✓ | ✓ | ✓ | ✓ | ✓ | ✓ | ✓ | ✓ |
| **ACALOS** | 1_L15 | ✓ | ✓ | | | ✓ | ✓ | | | ✓ |
| | 1_L35 | ✓ | ✓ | | | ✓ | ✓ | | | ✓ |
| | 1_diff | ✓ | ✓ | | | ✓ | ✓ | | | ✓ |
| | 4_L15 | ✓ | ✓ | | | ✓ | ✓ | | | ✓ |
| | 4_L35 | ✓ | ✓ | | | ✓ | ✓ | | | ✓ |
| | 4_diff | ✓ | ✓ | | | ✓ | ✓ | | | ✓ |

### II.3.2 Classification model

In (Saak et al., 2022), a random forest (RF, (Breiman, 2001) classification model was built to classify new patients into the profiles, which resulted in adequate performance (precision and sensitivity > 0.75 for the majority of the profiles). RF models work well with smaller sample sizes and provide an inherent feature importance index. We, therefore, also selected RF for the current analyses. RF uses an ensemble of trees, where the results of independent trees are aggregated, and the most frequently predicted label is used as the final prediction. Within each tree, only a subset of the features is used to split the predictor space into smaller regions, which effectively decorrelates the trees.

To build the RF model we used the one-vs-all (OVA) design, as ensemble methods have shown good performance for multiclass problems (Galar et al., 2011), OVA provided good performance in Saak et al. (2022), and it eases the interpretation with respect to feature importance. The OVA design splits a multiclass model into k binary models, where k is equal to the number of profiles. That means each model predicts whether an instance belongs to a specific profile or not. The feature importance therefore always highlights features that are most important for distinguishing the profile of interest from all remaining profiles. Data imbalance naturally exists with OVA design. To counter the imbalance, we upsampled labels using Gaussian noise to the average amount of labels available for each profile, which means all profiles have at least the average amount of patients in each profile, while profiles with patient numbers above the average retained their larger sample size. The rationale behind this was that we wanted to keep a balance between upsampled and original data. The remaining imbalance was addressed by using weights for the labels in the training. In that way, mistakes for the label of interest are more costly in terms of prediction errors.

#### II.3.3 Train, validation, and test set

The complete dataset (global AP set) was split into a training (80%) and a test set (20%). The training set (containing 80% of the data) was then further split into a training (80%) and validation set (20%). For the training set we used 10 times repeated 10-fold cross-validation to get a better estimate of the prediction error. The validation set was then used to evaluate the performances of the models on cases that were not used in the model training. After the model was specified with the training and validation set, the final model performance was evaluated with the test set.

#### II.3.4 Classification performance evaluation

Each classification model aims at reducing the prediction error, which is quantified by a specified evaluation metric. Evaluation metrics have different properties, making them useful for different prediction problems. For instance, accuracy is an evaluation metric that is easily interpretable, but does not perform well for imbalanced classification performances. We chose Cohen’s kappa (Cohen, 1960) as the evaluation metric for two reasons. First, Cohen’s kappa takes imbalances into account, by comparing the accuracy to the baseline accuracy that could be achieved by chance. Second, it proved to be the best evaluation metric among three others (balanced accuracy, Area under the precision recall curve, F1-score) for the classification model for profile set A in (Saak et al., 2022). Hence, in the model training process, we used Cohen’s Kappa to tune the number of considered features at each split.

To evaluate the general performance of the trained classification models for each feature set, we used two distinct but complementary metrics, namely sensitivity and precision. Sensitivity describes the proportion of correctly classified cases for the class of interest, whereas precision describes the proportion of misclassifications for the class of interest. Both sensitivity and precision were compared across feature sets and further, for the overall profile classification and single profiles.

Finally, we build a dummy prediction model, which does not include any features and predicts profile labels based on stratified sampling, that is, it reflects the relative frequency of patients contained in different profiles. That way, we could estimate the benefit of our prediction models as compared to the baseline dummy model which represents classification performance by chance.

#### II.3.5 Feature importance of the classification model

To estimate the feature importance of the RF models we used the inherent feature importance metric, namely the gini importance or rather the mean gini decrease (Breiman, 2001). The mean gini decrease is used in the training process to estimate how well a feature can split the labels across nodes in the ensemble of trees. A good split results in pure nodes where no misclassification occurs, whereas a bad split does not aid in separating the labels for the classification and leads to “impure” nodes. The mean gini decrease estimates how much a feature decreases the impurity on average across all nodes in the ensemble of trees, where a feature can be used multiple times in the same tree.

Overall feature importance across all profiles and feature importance specific for each profile was calculated and compared via a feature importance plot. We transformed the mean gini decrease to percentages (with the total mean gini decrease of a profile equaling 100 %) to show the contribution of each feature for each profile separately.

## III. Results

### III.1 Generation of Profiles

For dataset A, the optimal profile number of 13 was obtained from (Saak et al., 2022). For dataset B, the optimal profile number was determined according to the profile generation pipeline described in the present paper. The distribution of estimated optimal profile numbers across the bootstrapped datasets can be found in Figure 3. Both 31 and 32 were most frequently selected across the bootstrapped datasets and there is only a marginal difference between these two profile sets in terms of frequencies. In contrast, estimated profile numbers higher or lower than 31/32 were selected less frequently for the bootstrapped datasets. Since profile number 31 was slightly more frequently estimated than profile number 32, we selected 31 profiles as optimal for dataset B. Just as in (Saak et al., 2022), the best covariance parameterization for dataset B was “VEI”. “VEI” refers to variable volume, equal shape, and coordinate axis orientation. It therefore allows clusters to be of different size but restricts them in terms of shape and axis alignment. Although the number of profiles may seem high, it can be partially attributed to the chosen clustering method. Fewer profiles could have sufficed in explaining the data, if larger datasets were available that would allow for more complex clustering structures (fewer covariance matrix restrictions). Conversely, by increasing the number of clusters, less complex clustering structures than the one used here could still capture the same underlying data structure (Bouveyron et al., 2019).

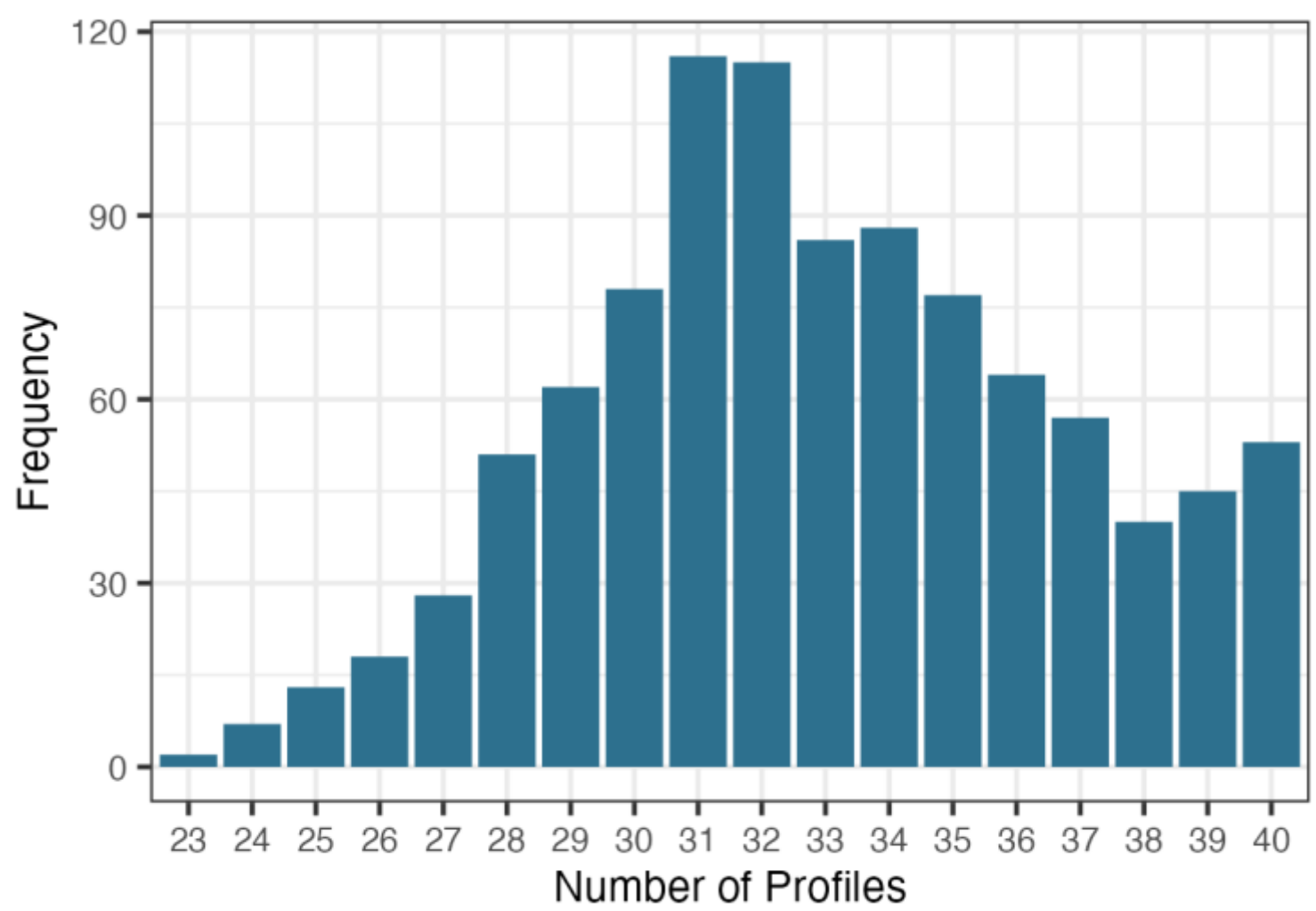

*Figure 3: Distribution of optimal profile numbers across the bootstrapped datasets*

### III.2 Merging Profiles

The profiles of the two datasets were merged using the mean overlapping density. This requires the selection of criteria to stop the merging process and result in the final profile set. Figure 4 displays the two criteria that we used to select the proposed combined profile set.

Figure 4A displays the highest overlap (i.e., profiles to be merged) next to the median overlap across all profiles for each iteration of the merging pipeline. Two cutoffs are depicted that show two potential profile sets. The first cutoff at 20 iterations is characterized by a steepening decrease of the highest overlap, next to a reduction in variance. The second cutoff at 32 iterations precedes an even steeper slope decrease and high variations in variance. The high variations in variance indicate that merges occurred where profile ranges became much broader such that the similarity between profiles could increase again. Since this is undesirable, a cutoff should occur prior to high variations in the variance.

Figure 4B displays the overlapping densities of the two profiles to be merged at each merging iteration. The mean overlap corresponds to the highest overlap from Figure 4A. The two cutoffs are indicated with the dashed green lines. We can observe a general decrease in overlap with increasing merging iteration, that corresponds to the results from Figure 4A but provide us with insights into which features drove the decrease in overlap. We can observe that especially after the cutoff at 32 iterations the mean overlap decreases substantially across multiple features, which indicates that profiles would be merged that can be distinguished. Hence, merging the profiles beyond 32 iterations would result in a substantial loss of information. This widespread decrease in overlap is not as pronounced for the cutoff at iteration 20. Hence, for the following analyses the profile set at 32 iterations was selected, which leads to a proposed combined profile set with thirteen APs. The corresponding analyses for the profile set at 20 iterations can be found in the supplementary material.

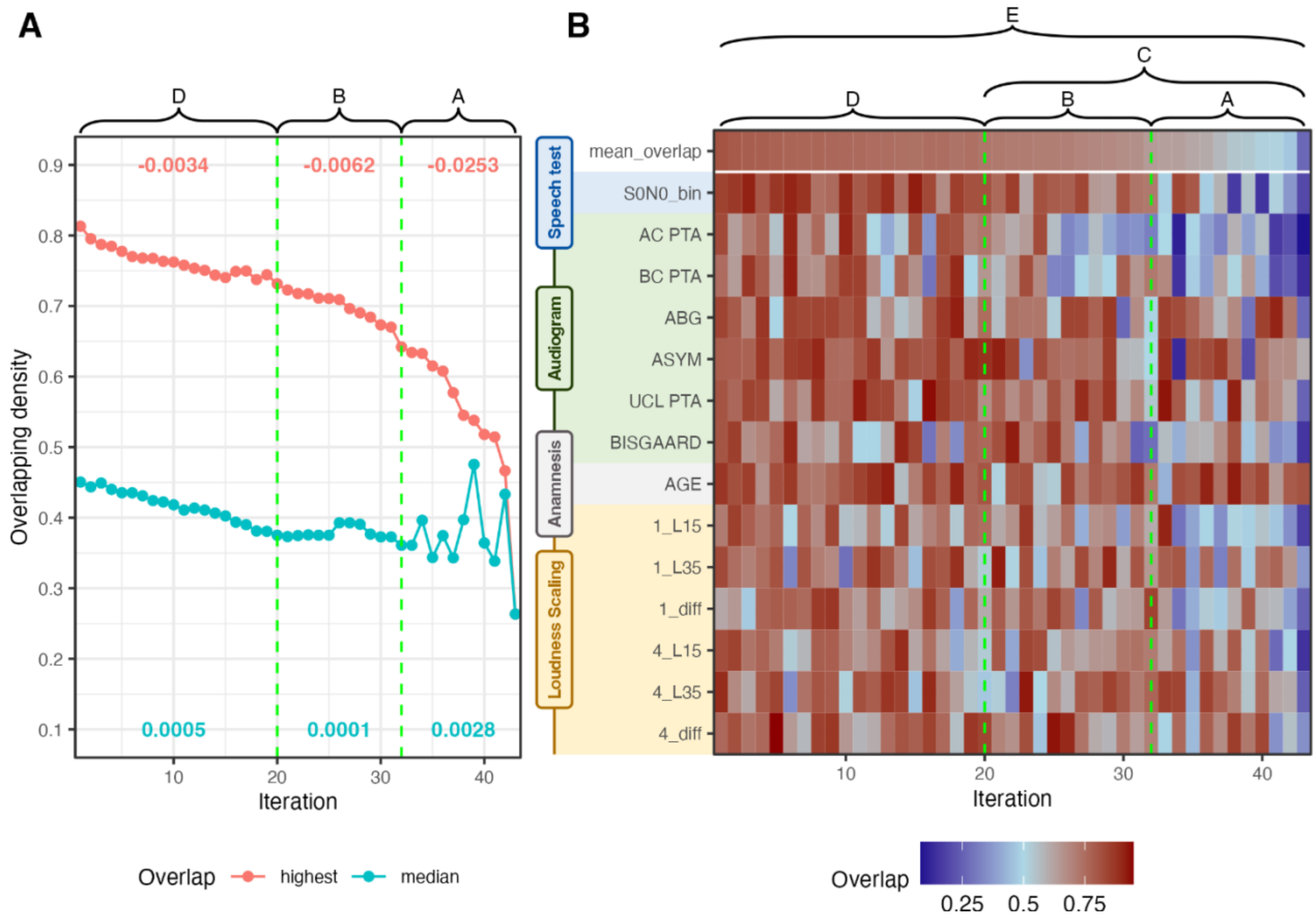

*Figure 4: Merging iterations. Vertical green lines indicate cutoff candidates (A) Median and highest overlapping density across the merge iterations. Red numbers indicate the slope of highest overlapping density for the respective merging area; turquois numbers indicate the variance of median overlapping density of the respective merging area.* ***(B)*** *Overlapping densities across features for the two profiles to be merged at the given iteration. A, B, C, D and E indicate different iteration sets.*

### 3. Feature importance of the merges

To ensure interpretability of the merging iterations, we estimated which features were most important for merging profiles. Features important for merging indicate they are less relevant for profile distinction, as they are similar across profiles. Conversely, features that are less important for merging can be interpreted as features relevant for profile separability. Hence, these features are determined as important features. Figure 5 displays the feature importance of the merging pipeline for different merging areas (A - E). A corresponds to the selected profile set at iteration 32; C to the profile set at iteration 20 (in more detail in the appendix); E visualizes profile importance across all iterations. B and D display feature importance for remainders of the iterations.

Feature importance, or profile separability for the 13 profiles of profile set 32 (merging area A) is mostly based on air- and bone-conduction PTA, the Bisgaard class, the SRT, and L15 and the difference of L35-L15 of the ACALOS for 1 kHz. For the complete set (E) the SRT becomes less important, and the L35-L15 difference becomes more relevant, next to L15 for 4 kHz. This occurs due to the higher importance of ACALOS features in merging area D. This means, that in the first profile merges in D (or later profile splits – if going from right to left iterations), loudness scaling based features are merged first (ACALOS information is lost), whereas in A, profile merging is driven more by audiogram-based features, the SRT, and L15 from the ACALOS.

Vice versa, this also implies that if we would start from only 2 profiles and would split until iteration 1, in later split iterations (profile sets with a higher number of profiles) more detail with respect to loudness scaling is added to the profiles.

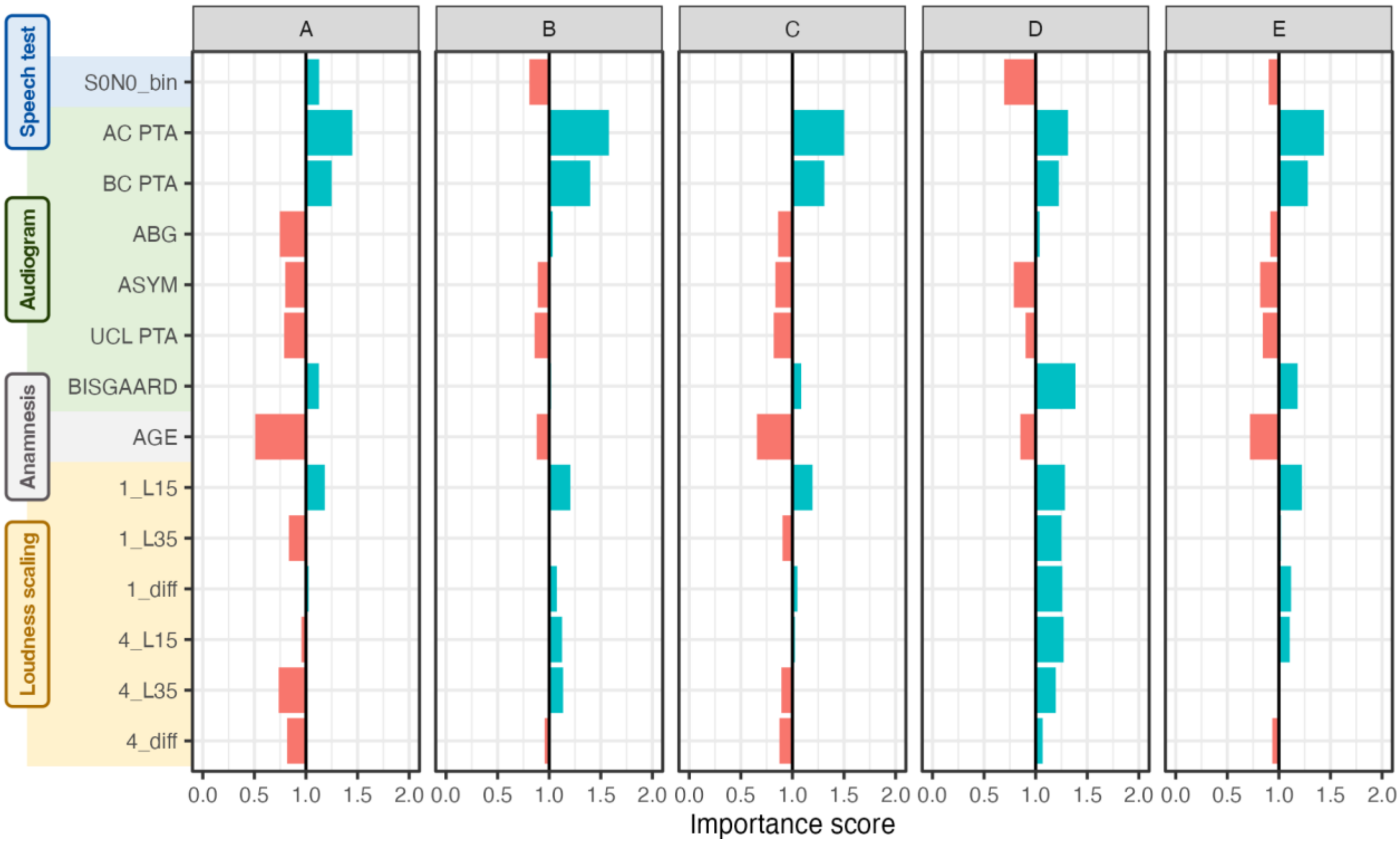

*Figure 5: Feature importance for different merging areas. A, B, C, D and E correspond to the indicated areas in Figure 4. Mean feature overlap was calculated for each feature in each merging area and then subtracted from 1 to indicate higher importance with higher scores. Subsequently, scores are divided by the mean overlap of all features of the merging set. Scores, thus, indicate the higher or lower importance than the average. Turquoise bars with values above 1 indicate higher importance; red bard with values below 1 indicate lower importance.*

**III.4 Proposed profile set**

Figure 6A visualizes the proposed profile set with 13 APs. Both the profile ranges for each feature (A) and a proposal for single profile visualization are depicted (B). We can observe distinct patterns across APs. Profile 13 corresponds to a normal hearing profile and profile 1 has the highest SRTs. In contrast, profile 6 shows the highest impairment in the PTA. Profile 8 contains the largest number of patients, corresponding to a mild high-frequency hearing loss, and profile 5 shows a similar pattern to profile 8, but with a higher degree of impairment. Overall, the profiles cover a large range of hearing deficits in terms of test measurement ranges. We can see a distinctiveness of the profiles based on the SRT, audiogram-related, and loudness scaling-related features. The age was not found important in the two datasets.

In the SRT range of -5 to 0 dB SNR differences between profiles are mainly driven by audiogram and loudness scaling based features. For instance, profiles 10 and 11 show similar SRTs, but differ regarding loudness perception and the AC PTA. Further, for profile 10 an asymmetry is present which could partly explain the higher AC PTAs. As expected, the presence of a higher asymmetry and a higher air-bone gap compensate for higher AC PTAs in terms of higher SRTs.

We can see a clear inverse trend of the dynamic ranges for the ACALOS 1 and 4 kHz to the SRT. That means, generally a higher SRT is accompanied by a reduced dynamic range, which is in line with existing research (Gieseler et al., 2017; Moore & Glasberg, 1993; Van Esch & Dreschler, 2015; Villchur, 1974). This trend fluctuates, when an ABG and asymmetry is also present in the profile.

The single profile visualization (Figure 6B) aids in visualizing the pattern for a single profile. Not all profiles are displayed, but the remainder can be found in the supplementary material. The polar plots depict the normalized median difference of each profile to the normal hearing profile (green circle). We can clearly see the impact that the presence of an ABG or asymmetry has on the relation between SRT and AC PTA (Profile 1 vs. Profile 3), that is, a smaller SRT results from a higher AC PTA in profile 3 due to asymmetry and/or ABG that both mitigate the general SRT deterioration with increasing AC PTA.

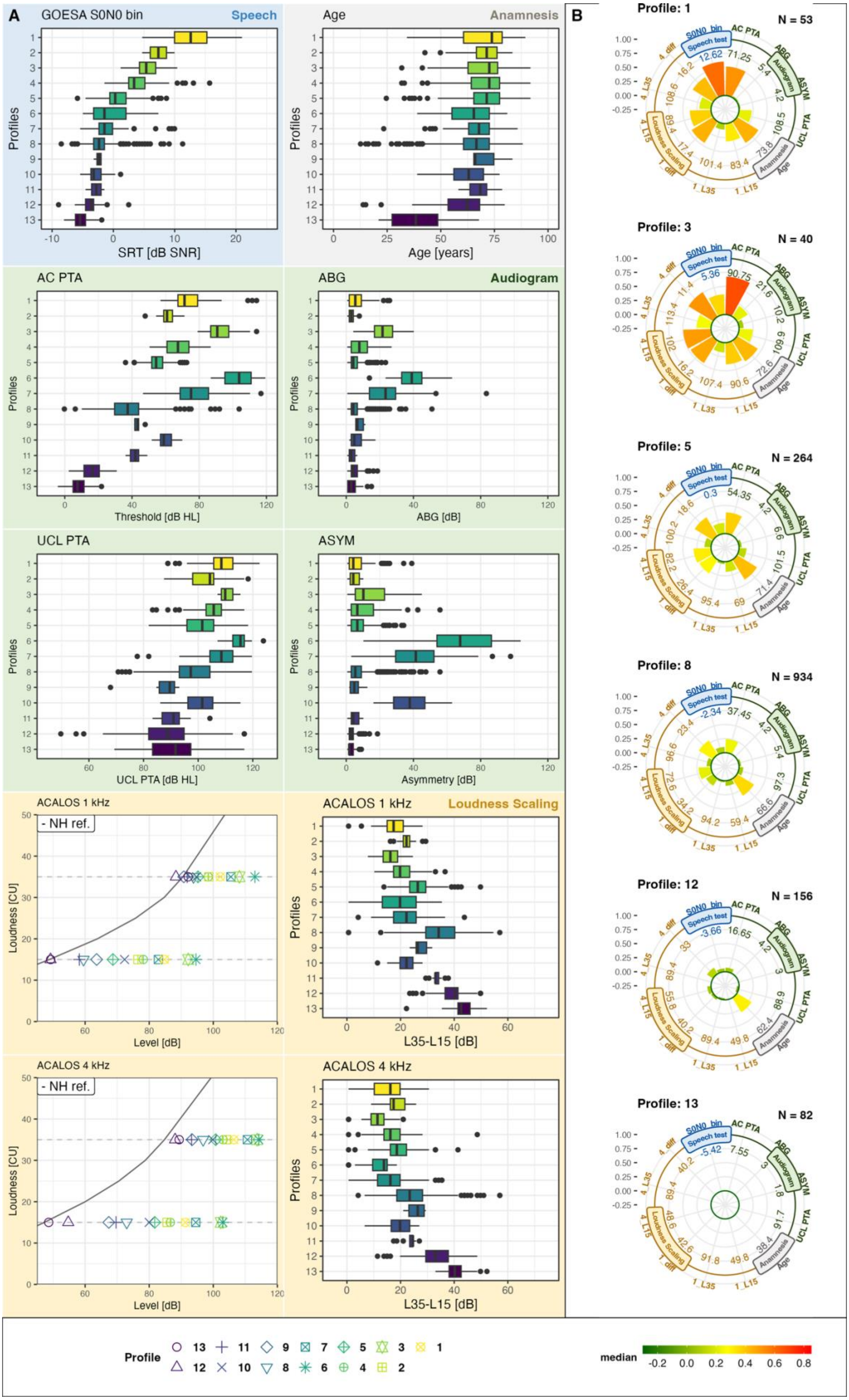
A
GOESA S0N0 bin
Speech
SRT [dB SNR]
Age
Anamnesis
Age [years]
AC PTA
Threshold [dB HL]
ABG
Audiogram
ABG [dB]
UCL PTA
UCL PTA [dB HL]
ASYM
Asymmetry [dB]
ACALOS 1 kHz
Loudness Scaling
- NH ref.
Loudness [CU]
Level [dB]
L35-L15 [dB]
ACALOS 4 kHz
Profiles
B
Profile: 1
N = 53
Profile: 3
N = 40
Profile: 5
N = 264
Profile: 8
N = 934
Profile: 12
N = 156
Profile: 13
N = 82
Speech test
Audiogram
Anamnesis
Loudness Scaling
Profile
median

*Figure 6: The 13 proposed APs across the speech, anamnesis, audiogram, and loudness scaling domain. (A) Profile ranges are depicted for all features and are ordered with respect to the increasing median SRT. (B) View of selected singular APs. The remainder of the singular APs can be found in the supplementary materials. Data is referenced to the NH profile (value 0). All bar values represent the median deviation from the NH profile, whereas the numbers indicate the true median value.*

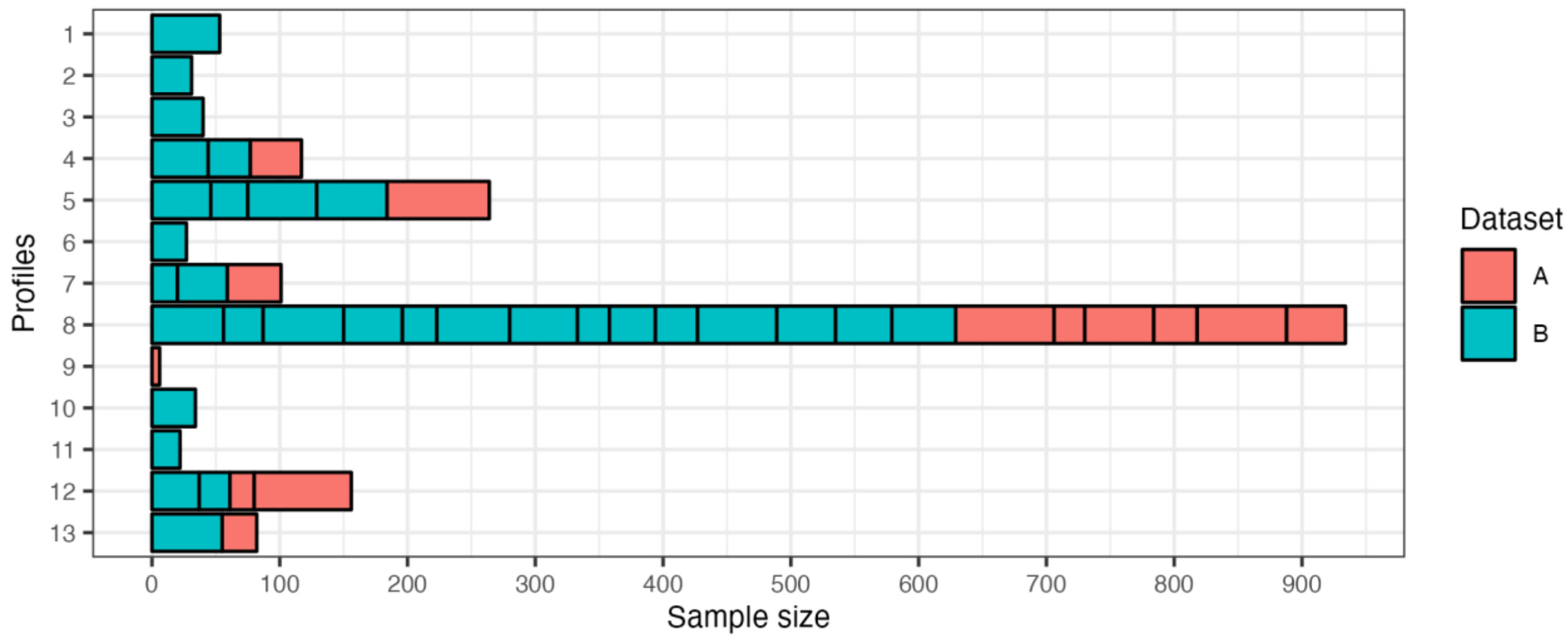

*Figure 7: Distribution of profiles that were merged to result in the selected 13 auditory profiles. A corresponds to dataset A and presents the previous 13 profiles detailed in (1). B refers to the new dataset B and the 31 profiles that were defined as optimal by the profile generation pipeline. The x-axis depicts the sample size for each sub-profile, as well as the proposed 13 profiles.*

Figure 7 depicts the distribution of profiles from the two profile sets (A & B) in the final 13 profiles. The 13 profiles from profile set A have been merged with the 31 profiles from profile set B. Hence, the previous profiles from (Saak et al., 2022) are also included within the new profile set. However, since dataset B contains patients with a larger variety and more severe deficits, the previous profiles are merged in favor of retaining a broader profile distribution with the new profile set. The number of patients per profile corresponds to the relative frequency of patients contained in the datasets. This is because model-based clustering, used for profile generation, does not impose any constraints on cluster size with the variable volume parameterization (“VEI”).

### III.4 Classification models

Classification performance across profiles for different feature sets are shown in Figure 8. The general classification performance is adequate for the “APP” and “ALL” feature sets (Figure 8A), with “APP” performing best among all feature sets. All feature sets performed better on average than the dummy model but for different profiles different benefits are achieved (Figure 8B). The higher performance of the dummy model for AP 8 can be explained by the larger sample size of this profile, which increases the chances of belonging to this specific AP. The “single” feature groups performed worst among the feature groups. This can, however, be expected, as less information is available to discriminate between APs with “single” sets. For the “single” feature group “AG” (audiogram), and for the “combined” feature group “AG ACALOS” achieved the best performance. The feature groups “SRT ACALOS” and “AG SRT” performed comparable on average and differences can be observed when comparing performances across profiles. For profiles with fewer deficits (higher profile number) “SRT ACALOS” could generally discriminate better than “AG SRT“, and vice versa for profiles with higher deficits (lower profile number). The feature sets “APP” and “ALL” both perform better than “HA” from the usecase feature group, which does not contain ACALOS information. The results demonstrate the importance of using audiological tests beyond the audiogram to adequately classify patients into APs. All three measures contribute to better discriminability into the distinct APs and the benefit of including ACALOS information for better discriminability is shown.

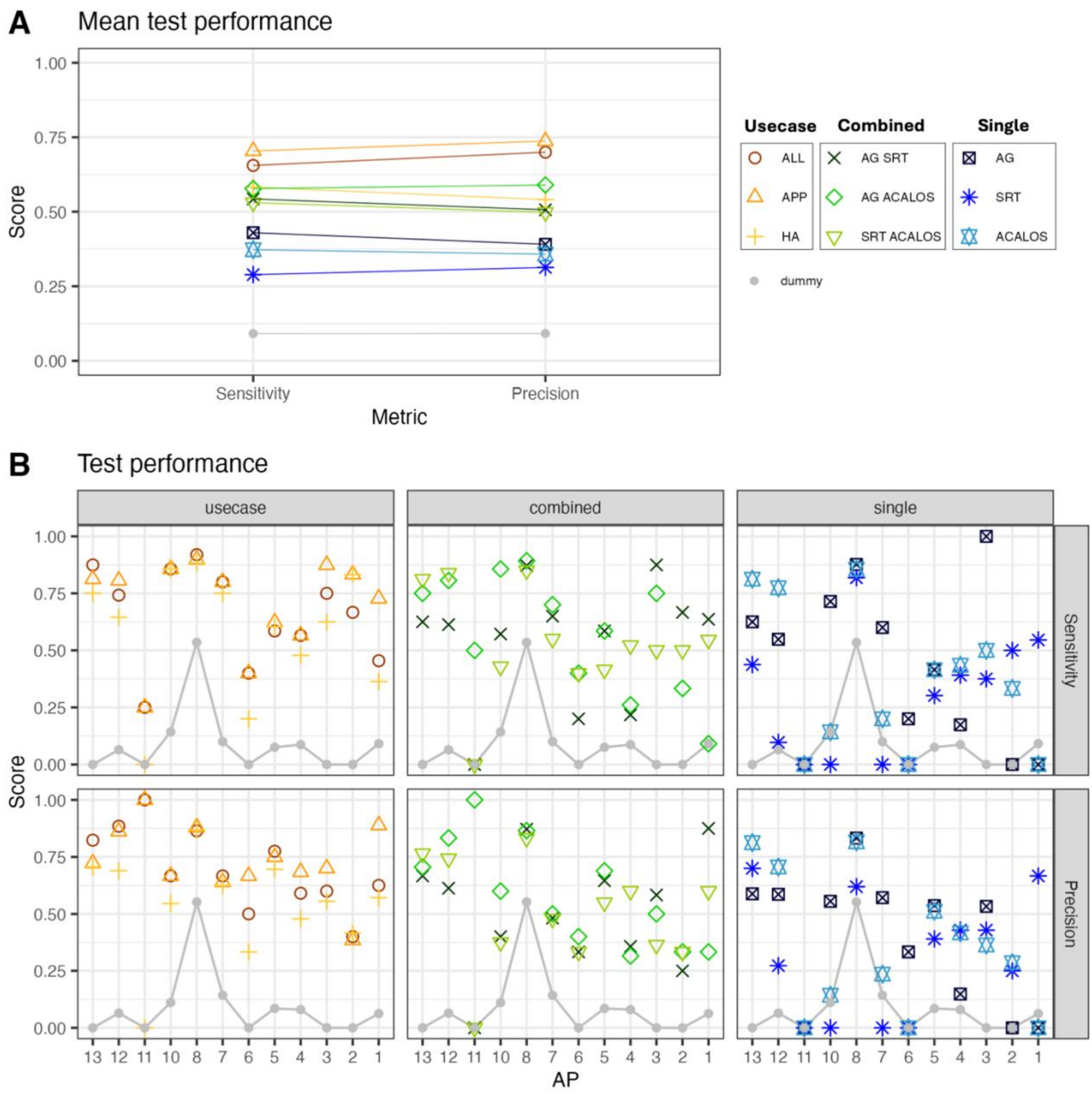

*Figure 8: Classification performance across profiles for different feature sets. Feature groups are categorized into "usecase", "combined", "single". Profile 9 is not displayed, as the sample size is too small for adequate training. (A) shows the mean test performance across profiles and (B) show the test performance for each profile. Dummy indicates the performance of the dummy model that predicts profile labels based on stratified sampling.*

To investigate the feature importance of the classification models, we selected the best-performing feature set, namely the "APP" feature set of the "use case" feature group (Figure 9). Hence, a reduced feature set performed best in classifying patients into the profiles. The AC PTA was determined to be overall most important for classifying profiles. More specifically, this means that it is most important for distinguishing between profiles in the presence of the remaining features. This is seconded by the SRT. We can, generally, observe that for profiles with higher SRTs (lower profile number) the SRT becomes more important. For instance, for distinguishing profile 1 and 2 from the remaining profiles, the SRT is more important than the AC PTA. We can further observe the most important features by comparing them to the mean importance of all features (Figure 9, dashed line in upper panel). The most important features are the SRT, AC PTA, asymmetry, and L15 for 1 kHz, and the importance of these features varies for different profiles. Profiles with lower SRTs (higher profile numbers) show a slight trend for higher importance of ACALOS features for differentiating the profiles from all remaining profiles.

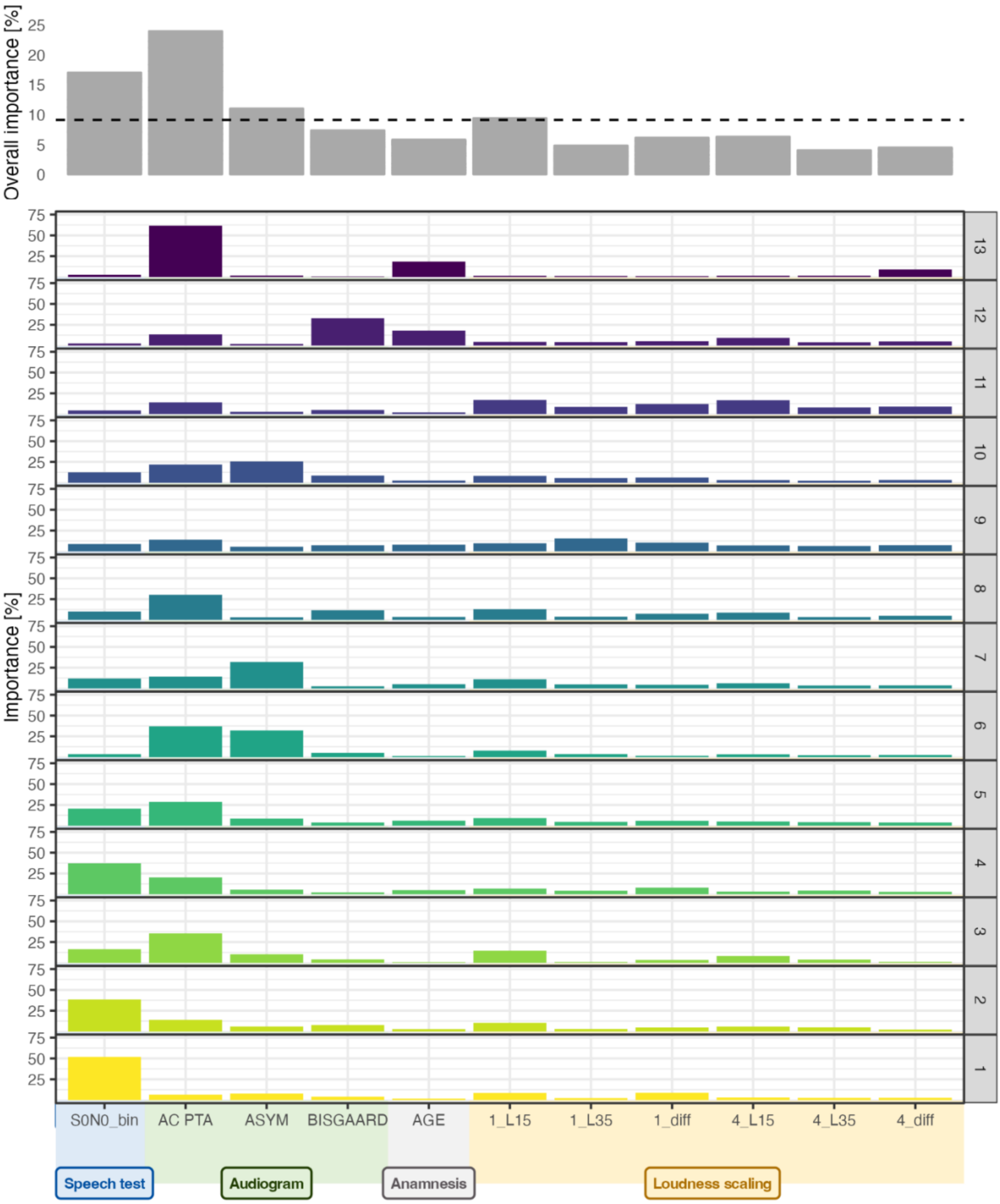

*Figure 9: Feature importance for "APP" feature set. Feature importance refers to percentage distribution of the mean gini decrease for each profile separately. Profile specific importance is shown by the colors. Overall importance (transformed to percentage) across all profiles is depicted by the grey bars. The dashed line indicates the mean percentage across features.*

## IV. Discussion

With the present study we aimed at extending the existing profiling approach such that datasets can be integrated into the auditory profiles (APs) towards an audiological patient population-based estimate of APs within a federated learning approach for profile merging. Our results show that APs generated across datasets can be plausibly merged using the mean density overlap of two profile distributions. The exact number of profiles is flexible and can be adjusted regarding the required detail of the profiles. We further trained classification models that allow for adequate classification of new patients into the APs using different feature combinations. Finally, we determined the importance of different features for both merging of and classification into the APs.

### IV.1 Auditory profile generation for dataset B

The optimal number of profiles generated for dataset B is 31. In comparison to the 13 APs from dataset A, this number can be considered rather high. However, one must consider the differences between the two datasets. First, dataset A is a research dataset where participants were recruited via a participant database of the Hörzentrum. In contrast, dataset B is a clinical dataset where participants approached the Hörzentrum themselves for the diagnostic support of an ENT physician. As a result, dataset B contains a larger sample covering a broader range of hearing loss including patient patterns with more severe hearing loss, as evident from the higher SRTs in profiles 1-3, which were not merged with profiles from dataset A. Second, the two datasets vary in terms of additionally included audiological measures. For dataset A, the most prominent additional features to the common features are cognitive measures. Dataset B, in contrast, contains additional features such as the S0N90 condition (noise from one side), both monaurally and binaurally, and information from the tympanogram, valsalva, and otoscopy. As a result, these additional features, in combination with the larger variation in patient patterns can explain why dataset B resulted in a higher optimal profile number (see Section "Role of common and additional features for profile generation, merging, and classification").

### IV.2 Proposed profile set

Our proposed iterative merging procedure, using the highest mean density overlap between two APs, respectively, enables the integration of different datasets via APs (RQ1.1). The separately generated profiles from the two datasets can, thus, be combined to describe both datasets together. The newly proposed and merged/combined profile set covers a wide range of hearing deficits and extends the range of deficits to the APs of profile set A, which are described in detail in (Saak et al., 2022). Both sensorineural and conductive/mixed hearing losses (APs 3, 6, and 7) are covered within the profiles. Integrating the two datasets (A and B) has extended the range of hearing deficits contained in the profiles as compared to the profile set A. This is evident from the new profiles with higher SRTs (Figure 7, APs 1-3) covering patients with more severe hearing deficits. The APs now compactly describe varying hearing deficits patterns.

APs 8, 5, and 12 are the largest profiles in terms of sample size, respectively. APs 8 and 5 both represent high-frequency hearing loss, which is a common type of hearing loss in audiology, but with different severity. More specifically, we observe a more pronounced reduction in the dynamic range at higher frequencies compared to lower frequencies (1_kHz diff vs. 4_kHz diff), which can aid in better speech intelligibility, as compared to a uniformly reduced dynamic range observed in AP 1. This trend is stronger for AP 8, which is in line with the generally more severe hearing loss patterns of AP 5. Since these types of hearing loss are common in audiology, this might explain the rather large proportion of patients that belong to these profiles. In other words, it can be expected that a frequent hearing loss pattern will contain more patients than a less frequent hearing loss pattern. However, the current analyses pipeline does not check whether the repeated merges could have led to profiles that have become too coarse. Hence, it would be important in the future to implement such a mechanism (for more details see next section).

AP 12 also contains a rather large proportion of the patients and corresponds to a profile with only mild hearing loss. Here it is interesting to note, that both UCL PTA and 1_L35 have lower scores than the normal hearing reference profile 13. This could be an indicator for the presence of recruitment. AP 1 and 3, generally, vary with respect to the impairments across included measures. AP 1 has a worse SRT score, but scores closer to the normal hearing reference for ACALOS and audiogram features compared to AP 3. This can be explained by the larger ABG present in AP 3. The general prevalence of an asymmetry and an ABG in the APs highlights the importance of including these patient types in further research.

**IV.3 Merging procedure**

The newly proposed profile set contains information from the two profile sets (A & B). The previously generated 13 APs (profile set A) are also represented in the dataset B (RQ1.2). They are both merged with profiles from dataset B, and with profiles from dataset A. That means, the profile set A is now represented by fewer and coarser profiles, while additional profiles with more severe hearing deficits were added to the new proposed profile set. This behavior can be explained by the continuous merges leading to a coarser profile separation, more severe cases being present in the dataset B, and that profiles from profile set A were more similar to themselves than profiles from profile set B.

The profiles, further, are flexible regarding their precise number of profiles following the merging. Depending on the use-case of the profiles, a more detailed or a coarser separation between profiles may be needed, which is why we decided on a flexible cutoff. Generally, selecting a cutoff in the later merging steps will lead to fewer profiles with broader ranges and a coarser separation. Conversely, selecting a cutoff in the earlier merging stages will lead to a higher number of profiles and allows to investigate smaller differences between profiles (RQ1.3). For instance, for screening purposes, a small number of profiles may suffice in broadly estimating mild, moderate, or severe hearing deficits. To achieve this, profiles could be merged further than shown in the current paper to result in fewer profiles. In contrast, a hearing care professional may need a more detailed separation of patients to potentially incorporate information from the profiles into the fitting procedure for hearing aids. Researchers may need an even more detailed separation of profiles to investigate relations and effects of certain audiological features. Here, a cutoff could be selected based on the loss of information for the feature of interest. For instance, if a lot of information regarding a certain feature is lost after a merge, one might select a cutoff prior to the information loss. An example for a larger profile set is shown in the Appendix (Figures A3 and A4).

The similarity score develops plausibly in a continuous way across merging iterations and features can be identified that drive the AP merges. In that way the merging procedure allows to plausibly combine, compare, and characterize the content of two datasets containing common features. With the current cutoff procedure, we aimed at first demonstrating the general feasibility and plausibility of the profile merges, which was confirmed with the feature importance analyses of the merges.

However, for an automized and unbiased estimation of the global profile set a calculated cutoff procedure will be necessary in the future. While a variable cutoff can make profiles more flexible for different use cases, a global auditory profile requires a fixed definition. One possible approach would be to analyze the within-profile similarity to the between-profile similarity. Within-profile similarity refers to how similar patients are to each other within a profile, whereas between-profile similarity describes the similarity of patients between two separate profiles. The within-profile similarity could be estimated via bootstrapping. One can assume that the within-profile similarity will be higher than the between-profile similarity, as patients within a profile are likely to be more similar to each other than to patients in other profiles. Therefore, the midpoint between the calculated within-profile and between-profile similarity could be used to define the merging cutoff in the future.

With the future integration of additional datasets, there is a risk of profiles becoming too coarse due to the repeated merges. To mitigate this, a monitoring mechanism is needed to detect when a profile has become too coarse. One option here would be to identify potential sub-profiles within a profile that may have become too coarse via clustering models. Another option would be to check whether the within-profile similarity deviates from the expected range. If the similarity deviates, the profiles could be split into two related sub-profiles.

**IV.4 Classification model and its applications**

With the "APP" feature set, we achieve an adequate classification into the 13 combined APs. The APs could, therefore, be predicted with a combination of audiogram, ACALOS, age, and speech test information. The majority of features were determined important, with the most important features being the SRT, AC PTA, ASYM, and L15 1 kHz of the ACALOS (RQ2). As no bone conduction measures are included in this feature set, the current set of APs could also be measured via smartphone. Classification into APs could, thus, be performed on data collected via smartphones. For audiogram-based measures, a variety of implementations already exist (Chen et al., 2021), while implemented speech tests include, for instance, DIN tests (Van Den Borre et al., 2021), word recognition tests (Van Zyl et al., 2018), and the matrix sentence test (Kollmeier et al., 2015; Saak et al., 2024). As speech tests differ, it would be necessary to consider appropriate ways how to achieve comparability between the available speech tests in different datasets before merging APs generated on the respective dataset.

Hearing care professionals currently measure the audiogram for hearing aid fitting. Depending on the respective regulations and tests available in each country, a speech test in quiet or noise is also used for hearing aid indication. (Hoppe & Hesse, 2017). The manufacturer first-fit of a hearing aid is, however, only based on the audiogram. Speech or loudness measurements are not considered, which appears to be insufficient to cover all aspects involved for compensating a hearing loss (Kollmeier & Kiessling, 2018). This is also reflected in the features required to classify a patient with respect to the AP found here: While to some extent the classification into profiles also works with the "AG SRT" feature set, the benefit of including loudness scaling shows with the performance improvement of the "APP" feature set. Single measures (single feature group, such as, e.g., the audiogram), in contrast, did not perform well in classifying patients into the profiles. This demonstrates the inability of single measures to characterize the complete extent of hearing deficits sufficiently and shows that, in practice, a combination of measures is needed to adequately characterize audiological patients.

**IV.5 Overall feature importance and interpretability**

The feature importance of the merging procedure and the classification models are generally comparable and appear audiologically plausible (RQ3). For both merging and classification, the same common features were initially considered. For merging, we performed no further feature selection. However, for the classification models, the best performing model was selected. Consequently, only features from the best performing model "APP" were considered from the common feature set. This resulted in the AC PTA, SRT, and the L15 (1 kHz) of the ACALOS to be among the most important features. The most important features, thus, cover the combination of threshold information, speech intelligibility, and loudness perception for soft sounds at 1 kHz. Especially in the later merging iterations (Figure 4B & Figure 5 - merging area A), the speech intelligibility gains importance, which demonstrates the relevance of speech information for our combined profile set. We, therefore, conclude that profiles were merged plausibly, and the underlying procedure is explainable, as we can observe which feature information is lost in each merging iteration and which features are relevant. By investigating the feature importance of the merging procedure, we can observe, for instance, that some information of loudness scaling is lost in earlier iterations. This becomes evident from the higher feature importance for loudness scaling in merging area D (see Figure 5). To capture all differences regarding loudness scaling, an earlier cutoff could be selected.

Our feature importance results are in line with existing results that highlight the importance of characterizing hearing deficits beyond the audiogram (Musiek et al., 2017). While the audiogram is often seen as the gold standard for characterizing hearing loss, it cannot characterize every aspect of existing deficits (Gieseler et al., 2017; Musiek et al., 2017; Sanchez-Lopez et al., 2021; Van Esch & Dreschler, 2015). Instead, a combination of threshold- and suprathreshold-based methods is needed, which are covered by audiogram, loudness scaling and speech intelligibility in the present research. Loudness scaling provides additional information beyond the audiogram (Kollmeier & Kiessling, 2018; Launer et al., 1996; Oetting et al., 2016; Van Beurden et al., 2021), which is confirmed by the better performance of the classification models if ACALOS is included ("HA" vs. "ALL" and "APP"). This effect is present even if the UCL PTA of the audiogram is included, hence proving the benefit of ACALOS beyond the UCL measure of the audiogram. One interesting finding is that the "AG SRT" and "SRT ACALOS" classification models perform comparable on average. However, profiles with lower SRTs (better speech intelligibility) are slightly better predicted by the SRT in combination with loudness information, and profiles with higher SRTs (worse speech intelligibility) are better predicted by the combination of audiogram information with the SRT. This trend can also be observed in Figure 9, where there is a slight trend for higher importance of ACALOS in profiles with lower SRTs as compared to profiles with higher SRTs. One potential explanation could be that L15 from the ACALOS is related to the threshold of the audiogram, as it measures sounds that were perceived as soft. In that way, it could partly cover audiogram-related information and additionally cover loudness-based information. Here, we can note that especially 4 kHz diff (L35-L15, measure of the dynamic range) is more important for profiles with lower SRTs than for profiles with higher SRTs. This is plausible, as it can be expected that individuals that are close to the normal hearing reference have a higher dynamic range and this dynamic range decreases more rapidly with increasing hearing loss for higher frequencies. Regardless, the combination of SRT and ACALOS with audiogram-related information remains necessary for adequate classification performance. Hence, it appears crucial to include threshold and suprathreshold information in characterizing hearing deficits.

**IV.6 Role of common and additional features for profile generation and merging**

The available features vary between datasets, and consequently also between profile generation and merging steps. In the profile generation steps, common features (see Table 1) and additional features are available (see descriptions of the datasets). In the merging step, profiles can only be merged based on common features. While we only show the common features in the current study, the additional features, such as S0N90 (noise from one side) are also available in the profiles. Although not all patients in a profile have data on these features, they enable us to estimate conditional probabilities of features ranges given a profile, thereby preserving information about the additional features in a descriptive manner. Merged profiles, therefore, contain integrated information on the common features used in the merging process. Note that both datasets employed included information on the individual audiogram, speech recognition in noise and loudness scaling which has some impact on the resulting finding, that all three information areas are relevant for classifying the individual patient. This would not have shown up if clinical datasets would have been employed with much less suprathreshold audiological information, which is both a strength and a potential pitfall of the current study. Additional features are not used to merge the APs, as they are not contained within both datasets, but were included for the generation of the profiles. The rationale behind this is that we aimed to make the first patient grouping based on the most informed choice using all available information - which includes the information contained in the additional features. This is beneficial for two reasons. First, the additional features can impact the initial grouping of the patients by allowing for finer distinctions, as compared to profile generation based solely on the common features. For example, using only the AC PTA and the SRT would create coarser groupings and miss certain subgroups that are revealed when including features from the ACALOS. Likewise, it can be assumed that relevant additional features would have a similar impact on the profile groupings. For that reason, we have always generated the profiles on the largest available feature set. It follows, that a certain number of common features should be available to adequately merge profiles generated from different datasets and to classify patients into a given profile. If we combined the two datasets based on the common features prior to clustering, the overarching hearing loss patterns would remain similar, but certain differences would exist due to the exclusion of the additional features. These differences could be attributed to both the impact of the additional features and the merging procedure in general. To better understand the impact of additional features and to exploit the contained information and knowledge, it is essential to investigate their precise impact on the profile generation in future studies. This would allow us to estimate the relevance of these features for patient characterization and disentangle their impact from the merging procedure.

Second, while merges are limited to common features, it may be of interest to derive profile sets that contain distinct feature sets. For example, if dataset *X* includes additional information on spectro-temporal modulation thresholds, a specific profile set could be generated by merging the profiles from dataset X only with profiles from datasets that also contain this information. These combined profiles could serve as sub-profile sets, which could also be merged with the global profile set based on the common measures. Hence, by always generating the profiles on the largest available feature set, rather than just common measures, the resulting profile sets are more flexible and can be merged with other datasets that share different features. Regardless, given the provided feature importance analysis, we advocate for using a combination of threshold, loudness- and speech test information. That is because these measures were consistently estimated as being important for both the merging and the later classification into the profile. However, depending on the use case and data availability, profile generation and potential later merging may also be used on an exploratory basis to learn about the content of a dataset at hand.

Another property of the additional features is that they could be used to infer probabilities within a given profile where data for the additional profiles is available. Here, it could be of interest to investigate whether certain feature ranges occur more frequently in certain profiles – especially if profiles were merged with at least one further profile containing the additional profile.

**IV.7 Towards an estimate of the audiological patient population via combined auditory profiles**

The combined profile set is the next step towards a better estimate of an audiological patient population-based AP set. Here, the audiological patient population primarily focuses on patients that have actively engaged with their own hearing loss by either visiting research institutes, clinics, or hearing care professionals. In the future, additional datasets need to be integrated to cover further hearing deficits, as well as information from different audiological measures. Different audiological measures not included in the current global auditory profile set, could be reported in a sub-profile set that combines multiple profile sets sharing common measures. This sub-profile set can complement the global profile set, providing more detailed estimates for specific measures of interest that are not included in the global profile set. Generally, it would be especially valuable if datasets were acquired via stratified sampling to improve the estimate of an audiological patient population-based AP set. While the new set of APs includes new information, continuing to integrate additional datasets can improve profile definitions and classification models. Currently, one large profile (AP 8) describes most of the patients due to the obviously high relative frequency of this profile (corresponding to a moderate high-frequency hearing loss, potentially age-related hearing loss) in the two employed datasets. Therefore, including additional datasets can help in better defining the remaining profiles based on a higher, more representative number of patients. If representative data for the whole population were included, the relative frequencies of patients contained in different profiles would allow for a prevalence estimate of different profiles. In addition, the current datasets mainly cover hearing aid candidates, whereas cochlear implant candidates are rarely included. An audiological patient population-based set should, however, include all degrees of hearing deficits and be as complete as possible. For this, data standards will be important. If data formats and required meta data are standardized across institutions, barriers for integrating datasets are reduced and could pave the way for big data analyses in the field of audiology.

The AP merging approach presented in this study offers a federated learning framework that, when additional datasets are included, can be used to obtain an AP set representative of the audiological patient population, without having to share individual patient data. The federated learning procedure includes the steps to generate and later merge profiles. One important property of the merging step is its ability to work with anonymized data. This is important to integrate datasets containing sensitive clinical data which often underly data sharing restriction (Chassang, 2017). For these datasets, the profile generation could occur at the sensitive data location and only the anonymized count data would be shared. As overlapping density is calculated for each feature separately and then averaged, it also enables shuffling patient records for each feature to completely anonymize the patient data. This does not impact the distributional descriptions of each feature contained in the profiles and therefore enables integration of sensitive datasets to work towards an audiological patient population-based estimate of APs.

To reach an audiological patient population-based estimate of APs and provide a fully functioning federated learning framework several further steps should be taken. First, to reach such a patient population-based estimate, we need to continue to merge profiles generated on different datasets until profiles converge on a final set of APs. A convergence would mean, that no new APs are added to the existing set of APs when integrating additional datasets. This follows the common principle of convergence in optimization algorithms, where the optimal solution (here sets of APs) is selected when parameters converge (Hastie et al., 2009). To achieve this, a fixed calculated cutoff would be needed, such that one can estimate whether profiles already converged, or a new profile should be added to the existing global profile set. Hence, future studies should investigate how to best define a fixed cutoff. Second, the precise impact of additional features on the generation of profiles should be disentangled from the merging procedure, such that we can pinpoint the impact these features have and evaluate their value for a good characterization of the audiological patient population. Third, a procedure should be developed that evaluates if profiles have become too coarse due to repeated merges and would thus warrant being split into smaller profiles again. This could make the merging procedure more robust and ensure good separability of the profiles.

In the future, with a sufficiently large AP set, in terms of included patients and varying deficits, it could be interesting to investigate the feasibility of profile-based first-fits for hearing aids. Profile-based fits have the potential to reduce the time required to fine-tune hearing aids, by incorporating information from the audiogram, speech test results, and loudness scaling. Especially for the individual selection of parameters for additional signal processing parameters a manufacturer-independent profile-based recommendation would help hearing care professionals to better individualize the first-fit parameters. Furthermore, first-fits could already include information on both thresholds and loudness perception. The benefit of using loudness-based fitting has been shown by Kramer et al. (2020) and Oetting et al. (2018). Since they highlight the importance of binaural broadband loudness scaling, future profile generation and evaluation should also include binaural broadband categorical loudness scaling.

## V. Conclusions

The current study demonstrates the feasibility of integrating datasets using the proposed profile generation and merging procedure, which together form a one-shot federated learning approach for a combined characterization of the content of audiological datasets. Combining the two datasets yields a new combined set of APs, which consists of 13 APs. Profiles can generally be well characterized based on the three dimensions: audiogram-, speech test-, and loudness scaling features.

We further enable the classification of patients into the APs using random forest classification models. The best performing classification models include a combination of these three measures (audiogram, speech test, loudness scaling), excluding the bone-conduction audiogram. While classification models tailored for hearing care professionals, which use only the audiogram and a speech test, are also available, their performance can be improved if loudness scaling is included.

Audiogram-, speech test-, and loudness scaling-based measures provide complementary information that aid in characterizing patients and are consistently determined as important for both merging and classification. Even though this finding is based on the composition of the two underlying datasets that both include these measures, we nevertheless advocate for the inclusion of these measures for detailed patient characterization.

Towards an audiological patient population-based set of APs, it is necessary to incorporate additional datasets using the proposed method. These datasets should include a wider range of audiological subpopulations, such as cochlear implant users, thereby extending the APs contained. To make the merging procedure more robust, further research is needed to optimize the cutoff for the profiles and integrate a mechanism that evaluates the coarseness of the profiles due to repeated merges. The privacy-preserving approach, employing federated learning, which involves the potential to share only data distributions from locally generated profiles, facilitates the integration of datasets that are subject to privacy restrictions.

## VI. Conflict of interest

The authors declare that the research was conducted in the absence of any conflicts of interest.

## VII. Statements and Declarations

*Ethical Considerations*
Our institution does not require ethical approval when already existing datasets are analyzed.

*Informed Consent*
Not applicable.

## VIII. Funding

This work was supported by the Deutsche Forschungsgemeinschaft (DFG, German Research Foundation) under Germany's Excellence Strategy −EXC 2177/1 – Project ID 390895286. MB was funded by the Deutsche Forschungsgemeinschaft (DFG, German Research Foundation) – Projektnummer 496819293, and the "Fondation Pour l'Audition" (FPA IDA10).

## IX. Data availability

The data analyzed in this study was obtained from Hörzentrum Oldenburg gGmbH, the following licenses/restrictions apply: According to the Data Usage Agreement of the authors, the dataset B analyzed in this study can only be shared upon motivated request. Requests to access these datasets should be directed to DO, oetting@hz-ol.de. Dataset A is publicly available and can be found here: https://doi.org/10.5281/zenodo.14177903. The analysis scripts can be found here: Zenodo, https://doi.org/10.5281/zenodo.13132817.

## X. References

Audigier, V., Husson, F., & Josse, J. (2016). A principal component method to impute missing values for mixed data. *Advances in Data Analysis and Classification*, *10*(1), Article 1. https://doi.org/10.1007/s11634-014-0195-1

Banerjee, A., & Shan, H. (2017). Model-Based Clustering. In C. Sammut & G. I. Webb (Eds.), *Encyclopedia of Machine Learning and Data Mining* (pp. 848–852). Springer US. https://doi.org/10.1007/978-1-4899-7687-1_554

Bisgaard, N., Vlaming, M. S. M. G., & Dahlquist, M. (2010). *Standard Audiograms for the IEC 60118-15 Measurement Procedure*. https://journals.sagepub.com/doi/abs/10.1177/1084713810379609

Bouveyron, C., Celeux, G., Murphy, T. B., & Raftery, A. E. (2019). *Model-Based Clustering and Classification for Data Science: With Applications in R*. Cambridge University Press.

Brand, T., & Hohmann, V. (2002). An adaptive procedure for categorical loudness scaling. *The Journal of the Acoustical Society of America*, *112*(4), 1597–1604. https://doi.org/10.1121/1.1502902

Breiman, L. (2001). Random Forests. *Machine Learning*, *45*(1), 5–32. https://doi.org/10.1023/A:1010933404324

Buuren, S. van, & Groothuis-Oudshoorn, K. (2011). mice: Multivariate Imputation by Chained Equations in R. *Journal of Statistical Software*, *45*, 1–67. https://doi.org/10.18637/jss.v045.i03

Chassang, G. (2017). The impact of the EU general data protection regulation on scientific research. *Ecancermedicalscience*, *11*, 709. https://doi.org/10.3332/ecancer.2017.709

Chen, C.-H., Lin, H.-Y. H., Wang, M.-C., Chu, Y.-C., Chang, C.-Y., Huang, C.-Y., & Cheng, Y.-F. (2021). Diagnostic Accuracy of Smartphone-Based Audiometry for Hearing Loss Detection: Meta-analysis. *JMIR mHealth and uHealth*, *9*(9), e28378. https://doi.org/10.2196/28378

Cohen, J. (1960). A Coefficient of Agreement for Nominal Scales. *Educational and Psychological Measurement*, *20*(1), 37–46.

De Sousa, K. C., Smits, C., Moore, D. R., Myburgh, H. C., & Swanepoel, D. W. (2022). Diotic and antiphasic digits-in-noise testing as a hearing screening and triage tool to classify type of hearing loss. *Ear and Hearing*, *43*(3), 1037–1048. https://doi.org/10.1097/AUD.0000000000001160

do Carmo, L. C., da Silveira, J. A. M., Marone, S. A. M., D’Ottaviano, F. G., Zagati, L. L., & von Söhsten Lins, E. M. D. (2008). Audiological study of an elderly brazilian population.

*Brazilian Journal of Otorhinolaryngology*, *74*(3), 342–349. https://doi.org/10.1016/S1808-8694(15)30566-8

Duan, R., Xiong, X., Liu, J., Liao, K. P., & Cai, T. (2024). *Federated One-Shot Ensemble Clustering* (arXiv:2409.08396). arXiv. http://arxiv.org/abs/2409.08396

European Commission. Directorate General for Health and Food Safety., Gesundheit Österreich Forschungs und Planungs GmbH., & Sogeti. (2016). *Study on Big Data in public health, telemedicine and healthcare: Final report.* Publications Office. https://data.europa.eu/doi/10.2875/734795

European Parliament & European Council. (2016). Regulation (EU) 2016/679 of the European Parliament and of the Council, of 27 April 2016 on the protection of natural persons with regard to the processing of personal data and on the free movement of such data, and repealing Directive 95/46/EC (General Data Protection Regulation). *Official Journal of the European Union*, *L 119*, 1–88.

Fraley, C., & Raftery, A. E. (2003). Enhanced Model-Based Clustering, Density Estimation,and Discriminant Analysis Software: MCLUST. *Journal of Classification*, *20*(2), 263–286. https://doi.org/10.1007/s00357-003-0015-3

Galar, M., Fernández, A., Barrenechea, E., Bustince, H., & Herrera, F. (2011). An overview of ensemble methods for binary classifiers in multi-class problems: Experimental study on one-vs-one and one-vs-all schemes. *Pattern Recognition*, *44*(8), 1761–1776. https://doi.org/10.1016/j.patcog.2011.01.017

Gemeinsamer Bundesausschuss. (2021). Richtlinie des Gemeinsamen Bundesausschusses über die verordnung von Hilfsmitteln in der vertragsärztlichen Versorgung. *Bundesanzeiger*, *B3*.

Gieseler, A., Tahden, M. A. S., Thiel, C. M., Wagener, K. C., Meis, M., & Colonius, H. (2017). Auditory and Non-Auditory Contributions for Unaided Speech Recognition in Noise as a Function of Hearing Aid Use. *Frontiers in Psychology*, *8*. https://www.frontiersin.org/articles/10.3389/fpsyg.2017.00219

Grant, R. W., McCloskey, J., Hatfield, M., Uratsu, C., Ralston, J. D., Bayliss, E., & Kennedy, C. J. (2020). Use of Latent Class Analysis and k-Means Clustering to Identify Complex Patient Profiles. *JAMA Network Open*, *3*(12), e2029068. https://doi.org/10.1001/jamanetworkopen.2020.29068

Guha, N., Talwalkar, A., & Smith, V. (2019). *One-Shot Federated Learning* (arXiv:1902.11175). arXiv. http://arxiv.org/abs/1902.11175

Hastie, T., Tibshirani, R., & Friedman, J. H. (2009). The elements of statistical learning: Data mining, inference, and prediction. *New York: Springer*, *2*, 1–758.

Hoppe, U., & Hesse, G. (2017). Hearing aids: Indications, technology, adaptation, and quality control. *GMS Current Topics in Otorhinolaryngology, Head and Neck Surgery*, *16*, Doc08. https://doi.org/10.3205/cto000147

Kalbe, E., Kessler, J., Calabrese, P., Smith, R., Passmore, A. P., Brand, M., & Bullock, R. (2004). DemTect: A new, sensitive cognitive screening test to support the diagnosis of mild cognitive impairment and early dementia. *International Journal of Geriatric Psychiatry*, *19*(2), 136–143. https://doi.org/10.1002/gps.1042

Kollmeier, B., & Kiessling, J. (2018). Functionality of hearing aids: State-of-the-art and future model-based solutions. *International Journal of Audiology*, *57*(sup3), S3–S28. https://doi.org/10.1080/14992027.2016.1256504

Kollmeier, B., Warzybok, A., Hochmuth, S., Zokoll, M. A., Uslar, V., Brand, T., & Wagener, K. C. (2015). The multilingual matrix test: Principles, applications, and comparison across languages: A review. *International Journal of Audiology*, *54 Suppl 2*, 3–16. https://doi.org/10.3109/14992027.2015.1020971

Kollmeier, B., & Wesselkamp, M. (1997). Development and evaluation of a German sentence test for objective and subjective speech intelligibility assessment. *The Journal of the Acoustical Society of America*, *102*(4), 2412–2421. https://doi.org/10.1121/1.419624

Konečný, J., McMahan, H. B., Yu, F. X., Richtárik, P., Suresh, A. T., & Bacon, D. (2017). *Federated Learning: Strategies for Improving Communication Efficiency* (arXiv:1610.05492). arXiv. https://doi.org/10.48550/arXiv.1610.05492

Kramer, F., Oetting, D., Schädler, M. R., Hohmann, V., & Warzybok, A. (2020). Speech recognition and loudness perception in normal-hearing and hearing-impaired listeners. *Forum Acusticum*, 3493–3493. https://doi.org/10.48465/fa.2020.0797

Launer, S., Holube, I., Hohmann, V., & Kollmeier, B. (1996). Categorical loudness scaling in hearing-impaired listeners Can loudness growth be predicted from the audiogram? *Audiological Acoustics*, *35*, 156–163.

Lê, S., Josse, J., & Husson, F. (2008). FactoMineR: An R Package for Multivariate Analysis. *Journal of Statistical Software*, *25*, 1–18. https://doi.org/10.18637/jss.v025.i01

McMahan, B., Moore, E., Ramage, D., Hampson, S., & Arcas, B. A. y. (2017). Communication-Efficient Learning of Deep Networks from Decentralized Data. *Proceedings of the 20th International Conference on Artificial Intelligence and Statistics*, 1273–1282. https://proceedings.mlr.press/v54/mcmahan17a.html

Moore, B. C., & Glasberg, B. R. (1993). Simulation of the effects of loudness recruitment and threshold elevation on the intelligibility of speech in quiet and in a background of speech. *The Journal of the Acoustical Society of America*, *94*(4), 2050–2062. https://doi.org/10.1121/1.407478

Musiek, F. E., Shinn, J., Chermak, G. D., & Bamiou, D.-E. (2017). Perspectives on the Pure-Tone Audiogram. *Journal of the American Academy of Audiology*, *28*(7), 655–671. https://doi.org/10.3766/jaaa.16061

Oetting, D., Hohmann, V., Appell, J.-E., Kollmeier, B., & Ewert, S. D. (2016). Spectral and binaural loudness summation for hearing-impaired listeners. *Hearing Research*, *335*, 179–192. https://doi.org/10.1016/j.heares.2016.03.010

Oetting, D., Hohmann, V., Appell, J.-E., Kollmeier, B., & Ewert, S. D. (2018). Restoring Perceived Loudness for Listeners With Hearing Loss. *Ear and Hearing*, *39*(4), Article 4. https://doi.org/10.1097/AUD.0000000000000521

Parimbelli, E., Marini, S., Sacchi, L., & Bellazzi, R. (2018). Patient similarity for precision medicine: A systematic review. *Journal of Biomedical Informatics*, *83*, 87–96. https://doi.org/10.1016/j.jbi.2018.06.001

Pastore, M., & Calcagnì, A. (2019). Measuring Distribution Similarities Between Samples: A Distribution-Free Overlapping Index. *Frontiers in Psychology*, *10*. https://doi.org/10.3389/fpsyg.2019.01089

Pfitzner, B., Steckhan, N., & Arnrich, B. (2021). Federated Learning in a Medical Context: A Systematic Literature Review. *ACM Transactions on Internet Technology (TOIT)*. https://doi.org/10.1145/3412357

Plomp, R. (1986). A signal-to-noise ratio model for the speech-reception threshold of the hearing impaired. *Journal of Speech and Hearing Research*, *29*(2), 146–154. https://doi.org/10.1044/jshr.2902.146

Roth, T. N., Hanebuth, D., & Probst, R. (2011). Prevalence of age-related hearing loss in Europe: A review. *European Archives of Oto-Rhino-Laryngology*, *268*(8), 1101–1107. https://doi.org/10.1007/s00405-011-1597-8

Saak, S., Huelsmeier, D., Kollmeier, B., & Buhl, M. (2022). A flexible data-driven audiological patient stratification method for deriving auditory profiles. *Frontiers in Neurology*, *13*. https://www.frontiersin.org/articles/10.3389/fneur.2022.959582

Saak, S., Kothe, A., Buhl, M., & Kollmeier, B. (2024). *Comparison of user interfaces for measuring the matrix sentence test on a smartphone* (arXiv:2401.17202). arXiv. https://doi.org/10.48550/arXiv.2401.17202

Sanchez-Lopez, R., Fereczkowski, M., Neher, T., Santurette, S., & Dau, T. (2020). Robust Data-Driven Auditory Profiling Towards Precision Audiology. *Trends in Hearing*, *24*, 2331216520973539. https://doi.org/10.1177/2331216520973539

Sanchez-Lopez, R., Nielsen, S. G., El-Haj-Ali, M., Neher, T., Dau, T., & Santurette, S. (2021). Auditory Tests for Characterizing Hearing Deficits in Listeners With Various Hearing Abilities: The BEAR Test Battery. *Frontiers in Neuroscience*, *15*. https://doi.org/10.3389/fnins.2021.724007

Schmidt, K. H., & Metzler, P. (1992). WST: Wortschatztest. *Beltz*.

Schwarz, G. (1978). Estimating the Dimension of a Model. *The Annals of Statistics*, *6*(2), 461–464. https://doi.org/10.1214/aos/1176344136

Sinkala, M., Mulder, N., & Martin, D. (2020). Machine Learning and Network Analyses Reveal Disease Subtypes of Pancreatic Cancer and their Molecular Characteristics. *Scientific Reports*, *10*(1), Article 1. https://doi.org/10.1038/s41598-020-58290-2

Stöver, T., Plontke, S. K., Guntinas-Lichius, O., Welkoborsky, H.-J., Zahnert, T., Delank, K. W., Deitmer, T., Esser, D., Dietz, A., Wienke, A., Loth, A., & Dazert, S. (2023). Struktur und Einrichtung des Deutschen Cochlea-Implantat-Registers (DCIR). *HNO*, *71*(12), 767–778. https://doi.org/10.1007/s00106-023-01309-7

Szatmari, T.-I., Petersen, M. K., Korzepa, M. J., & Giannetsos, T. (2020). Modelling Audiological Preferences using Federated Learning. *Adjunct Publication of the 28th ACM Conference on User Modeling, Adaptation and Personalization*, 187–190. https://doi.org/10.1145/3386392.3399560

Van Beurden, M., Boymans, M., Van Geleuken, M., Oetting, D., Kollmeier, B., & Dreschler, W. A. (2021). Uni- and bilateral spectral loudness summation and binaural loudness summation with loudness matching and categorical loudness scaling. *International Journal of Audiology*, *60*(5), 350–358. https://doi.org/10.1080/14992027.2020.1832263

Van Den Borre, E., Denys, S., Van Wieringen, A., & Wouters, J. (2021). The digit triplet test: A scoping review. *International Journal of Audiology*, *60*(12), 946–963. https://doi.org/10.1080/14992027.2021.1902579

Van Esch, T. E. M., & Dreschler, W. A. (2015). Relations Between the Intelligibility of Speech in Noise and Psychophysical Measures of Hearing Measured in Four Languages Using the Auditory Profile Test Battery. *Trends in Hearing*, *19*, 2331216515618902. https://doi.org/10.1177/2331216515618902

van Esch, T. E. M., Kollmeier, B., Vormann, M., Lyzenga, J., Houtgast, T., Hällgren, M., Larsby, B., Athalye, S. P., Lutman, M. E., & Dreschler, W. A. (2013). Evaluation of the preliminary auditory profile test battery in an international multi-centre study. *International Journal of Audiology*. https://www.tandfonline.com/doi/full/10.3109/14992027.2012.759665

Van Zyl, M., Swanepoel, D. W., & Myburgh, H. C. (2018). Modernising speech audiometry: Using a smartphone application to test word recognition. *International Journal of Audiology*, *57*(8), 561–569. https://doi.org/10.1080/14992027.2018.1463465

Villchur, E. (1974). Simulation of the effect of recruitment on loudness relationships in speech. *The Journal of the Acoustical Society of America*, *56*(5), 1601–1611. https://doi.org/10.1121/1.1903484

Xu, C., Schell-Majoor, L., & Kollmeier, B. (2024). *Development and verification of non-supervised smartphone-based methods for assessing pure-tone thresholds and loudness perception* (p. 2024.06.25.24309468). medRxiv. https://doi.org/10.1101/2024.06.25.24309468

Zokoll, M. A., Wagener, K. C., Brand, T., Buschermöhle, M., & Kollmeier, B. (2012). Internationally comparable screening tests for listening in noise in several European languages: The German digit triplet test as an optimization prototype. *International Journal of Audiology*. https://www.tandfonline.com/doi/full/10.3109/14992027.2012.690078